\documentclass[modern]{aastex62}

\usepackage{amsmath}
\usepackage{mathrsfs}
\usepackage{calc}
\usepackage{pgf}
\clearpage{}%
\clearpage{}%
\DeclareMathOperator*{\argmax}{argmax}
\newcommand{\fourier}[1]{\mathscr{F}\left\{#1\right\}}
\newcommand{\inversefourier}[1]{\mathscr{F}^{-1}\left\{#1\right\}}
\newcommand{\AR}{active region}
\newcommand{\dem}{$\mathrm{EM}(T)$}
\newcommand{\twait}[1][]{t_{\textup{wait}#1}}
\newcommand{\Tpeak}{T_\textup{peak}}

\defcitealias{barnes_understanding_2019-1}{Paper II}

\begin{document}

%%%%%%%%%%%%%%%%%%%%%%%%%%%%%%%%%%%%%%%%%%%%%%%%%%%%%%%%%%%%%%%%%%%%%%%%%%%%%%%
%                                   Title and Authors                         %
%%%%%%%%%%%%%%%%%%%%%%%%%%%%%%%%%%%%%%%%%%%%%%%%%%%%%%%%%%%%%%%%%%%%%%%%%%%%%%%
\title{Understanding Heating in Active Region Cores through Machine Learning I. Numerical Modeling and Predicted Observables}
\author[0000-0001-9642-6089]{W. T. Barnes}
\affiliation{Department of Physics \& Astronomy, Rice University, Houston, TX 77005-1827}
\affiliation{Lockheed Martin Solar \& Astrophysics Laboratory, Palo Alto, CA 94304}
\affiliation{Bay Area Environmental Research Institute, Moffett Field, CA 94952}
\author[0000-0002-3300-6041]{S. J. Bradshaw}
\affiliation{Department of Physics \& Astronomy, Rice University, Houston, TX 77005-1827}
\author[0000-0003-1692-1704]{N. M. Viall}
\affiliation{NASA Goddard Space Flight Center, Greenbelt, MD 20771}
\correspondingauthor{W. T. Barnes}
\email{barnes@lmsal.com}

%%%%%%%%%%%%%%%%%%%%%%%%%%%%%%%%%%%%%%%%%%%%%%%%%%%%%%%%%%%%%%%%%%%%%%%%%%%%%%%
%                                   Editorial Information                     %
%%%%%%%%%%%%%%%%%%%%%%%%%%%%%%%%%%%%%%%%%%%%%%%%%%%%%%%%%%%%%%%%%%%%%%%%%%%%%%%
\received{20 March 2019}
\revised{29 May 2019}
\accepted{7 June 2019}
\submitjournal{The Astrophysical Journal}

%%%%%%%%%%%%%%%%%%%%%%%%%%%%%%%%%%%%%%%%%%%%%%%%%%%%%%%%%%%%%%%%%%%%%%%%%%%%%%%
%                                   Abstract                                  %
%%%%%%%%%%%%%%%%%%%%%%%%%%%%%%%%%%%%%%%%%%%%%%%%%%%%%%%%%%%%%%%%%%%%%%%%%%%%%%%
\begin{abstract}
To adequately constrain the frequency of energy deposition in active region cores in the solar corona, systematic comparisons between detailed models and observational data are needed. In this paper, we describe a pipeline for forward modeling active region emission using magnetic field extrapolations and field-aligned hydrodynamic models. We use this pipeline to predict time-dependent emission from active region NOAA 1158 as observed by SDO/AIA for low-, intermediate-, and high-frequency nanoflares. In each pixel of our predicted multi-wavelength, time-dependent images, we compute two commonly-used diagnostics: the emission measure slope and the time lag. We find that signatures of the heating frequency persist in both of these diagnostics. In particular, our results show that the distribution of emission measure slopes narrows and the mean decreases with decreasing heating frequency and that the range of emission measure slopes is consistent with past observational and modeling work. Furthermore, we find that the time lag becomes increasingly spatially coherent with decreasing heating frequency while the distribution of time lags across the whole active region becomes more broad with increasing heating frequency. In a follow up paper, we train a random forest classifier on these predicted diagnostics and use this model to classify real AIA observations of NOAA 1158 in terms of the underlying heating frequency.
\end{abstract}
%% Keywords

%%%%%%%%%%%%%%%%%%%%%%%%%%%%%%%%%%%%%%%%%%%%%%%%%%%%%%%%%%%%%%%%%%%%%%%%%%%%%%%
%                                   Sections                                  %
%%%%%%%%%%%%%%%%%%%%%%%%%%%%%%%%%%%%%%%%%%%%%%%%%%%%%%%%%%%%%%%%%%%%%%%%%%%%%%%
%
%%%%%%%%%%%%%%%%%%%%%%%%%%%%%%%%%%%%%%%%%%%%%%%%%%%%%%%%%%%%%%%%%%%%%%%%%%%%%%%
%                                   Introduction                              %
%%%%%%%%%%%%%%%%%%%%%%%%%%%%%%%%%%%%%%%%%%%%%%%%%%%%%%%%%%%%%%%%%%%%%%%%%%%%%%%
\section{Introduction}\label{introduction}

Nanoflares\added{, energetic bursts of order $10^{24}$ erg resulting from small-scale reconnection \citep{parker_nanoflares_1988},} have long been used to explain the observed million-degree temperatures in the non-flaring solar corona. \deleted{Though originally pertaining to energetic bursts of order $10^{24}$ erg resulting from small-scale reconnection \citep{parker_nanoflares_1988}, the term \textit{nanoflare} is now synonymous with any impulsive energy release and is not specific to any particular physical mechanism \citep{klimchuk_key_2015}.} Due to their faint, transient nature, direct observations of nanoflares are made difficult by several factors, including inadequate spectral coverage of instruments, the efficiency of thermal conduction, and non-equilibrium ionization \citep{cargill_implications_1994,winebarger_defining_2012,barnes_inference_2016}. However, recent observations of ``very hot'' 8-10 MK plasma, the so-called ``smoking gun'' of nanoflares, have provided compelling evidence for their existence \citep[e.g.][]{brosius_pervasive_2014,caspi_new_2015,parenti_spectroscopy_2017,ishikawa_detection_2017}.

Critical to understanding the underlying heating mechanism is knowing whether the corona in non-flaring active regions is heated \textit{steadily} or \textit{impulsively}. \replaced{or more precisely}{However, because both waves and reconnection can lead to impulsive heating \citep{klimchuk_key_2015}, it is better to ask} at what frequency do nanoflares repeat on a given magnetic strand. In the case of low-frequency nanoflares, the time between consecutive events on a strand is long relative to its characteristic cooling time, giving the strand time to fully cool and drain before it is re-energized. In the high-frequency scenario, the time between events is short relative to the cooling time such that the strand is not allowed to fully cool before being heated again. Steady heating may be regarded as nanoflare heating in the very high-frequency limit. 

Before proceeding, we note that a magnetic \textit{strand}, the fundamental unit of the low-$\beta$ corona, is a flux tube oriented parallel to the magnetic field that is isothermal in the direction perpendicular to magnetic field. We make the distinction that a \textit{coronal loop} is an observationally-defined feature representing a field-aligned intensity enhancement relative to the surrounding diffuse emission, such that a single coronal loop may be composed of many thermally-isolated strands. Furthermore, we define the \AR{} \textit{core} as the area near the center of the \AR{} whose X-ray and EUV emission is dominated by closed loops with both footpoints rooted in the photosphere \added{within the \AR{}}.

In lieu of a direct observable signature of nanoflare heating, two parameters in particular have been used to diagnose the heating frequency in \AR{} cores: the emission measure slope and the time lag. These diagnostics provide \textit{indirect} signatures of the energy deposition via observations of the plasma cooling by thermal conduction, enthalpy, and radiation. We now discuss each of these observables in detail.

The emission measure distribution, $\mathrm{EM}(T)=\int\mathrm{d}h\,n_e^2$, where $n_e$ is the electron density and the integration is taken along the line of sight, is a useful diagnostic for parameterizing the frequency of energy deposition. Many observational and theoretical studies have suggested that the ``cool'' portion of the \dem{} (i.e. leftward of the peak, $10^{5.5}\lesssim T\lesssim10^{6.5}$ K), can be described by $\mathrm{EM}(T)\sim T^a$ \citep{jordan_structure_1976,cargill_implications_1994,cargill_nanoflare_2004}. The so-called \textit{emission measure slope}, $a$, is an important diagnostic for assessing how often a single strand may be reheated and has been used by several researchers to interpret \AR{} core observations in terms of both high- and low-frequency heating \citep[see Table 3 of][and references therein]{bradshaw_diagnosing_2012}. The ``cool'' emission measure slope typically falls in the range $2<a<5$, with shallower slopes indicative of low-frequency heating and steeper slopes associated with high-frequency heating. Many observational studies of active region cores have used the emission measure slope to make conclusions about the heating frequency \citep[e.g.][]{tripathi_emission_2011,warren_constraints_2011,winebarger_using_2011,schmelz_cold_2012,warren_systematic_2012,del_zanna_evolution_2015}.

To better understand observable properties of nanoflare heating, several researchers have used hydrodynamic models of coronal loops to examine how the emission measure slope varies with heating frequency \citep[e.g.][]{mulu-moore_can_2011,bradshaw_diagnosing_2012,reep_diagnosing_2013}. Most recently, \citet{cargill_active_2014} found that varying the time between consecutive heating events from 250 s (high-frequency heating) to 5000 s (low-frequency heating) could account for the wide observed distribution of emission measure slopes, with higher values of $a$ corresponding to higher heating frequency due to the \dem{} distribution becoming increasingly isothermal \citep[see also][]{barnes_inference_2016-1}.

In addition to the emission measure slope, the time lag analysis of \citet{viall_evidence_2012} has also been used by several workers to understand the frequency of energy release in \AR{} cores. The \textit{time lag} is the temporal delay which maximizes the cross-correlation between two time series, and, qualitatively, can be thought of as the amount of time which one signal must be shifted relative to another in order to achieve the best ``match'' between the two signals. As the plasma cools through the six EUV channels of the Atmospheric Imaging Assembly instrument \citep[AIA,][]{lemen_atmospheric_2012} onboard the Solar Dynamics Observatory spacecraft \citep[SDO,][]{pesnell_solar_2012}, we expect to see the intensity peak in successively cooler passbands of AIA according to the sensitivity of each channel in temperature space \citep{viall_patterns_2011}. Computing the time lag between light curves in different channels provides a proxy for the cooling time between channels and insight into the thermal evolution of the plasma. Calculating the time lag in each pixel of an AIA image can reveal large scale cooling patterns in coronal loops as well as the diffuse emission between loops across an entire \AR{}.

\citet{viall_evidence_2012} computed time lags for all possible AIA EUV channel pairs in every pixel of \AR{} NOAA 11082 and found positive time lags across the entire \AR{} core, indicative of cooling plasma. They interpreted these observations as being inconsistent with a steady heating model. \citet{viall_survey_2017} extended this analysis to the 15 active regions catalogued by \citet{warren_systematic_2012} and found overwhelmingly positive time lags, or cooling plasma, in all cases, with only a few isolated instances of negative time lags, or heating plasma. These observations are consistent with an impulsive heating scenario in which little emission is produced during the heating phase because of the time needed to fill the corona by chromospheric evaporation and the efficiency of thermal conduction. \citet{bradshaw_patterns_2016} predicted AIA intensities for a range of nanoflare heating frequencies in a model \AR{} and applied the time lag analysis to their simulated images. They found that aspects of both high- and intermediate-frequency nanoflares reproduced the observed time lag patterns, but neither model could fully account for the observational constraints, suggestive of a range of heating frequencies across the \AR{}. Additionally, \citet{lionello_can_2016} used a field-aligned hydrodynamic model to compute time lags for several loops in NOAA 11082 and concluded that an impulsive heating model could not account for the long ($>5000$ s) time lags calculated from observations by \citet{viall_evidence_2012}.

Any successful heating model must be able to reproduce the observed distribution of emission measure slopes and time lags. In order to carry out such a test, both advanced forward modeling and sophisticated comparisons to data are required. In this paper, we carry out a series of nanoflare heating simulations in order to better understand how the frequency of impulsive heating events on a given strand is related to observable properties of the plasma, notably the emission measure slope and the time lag as derived from AIA observations. To do this, we use a combination of magnetic field extrapolations, hydrodynamic models, and atomic data to produce simulated AIA emission which can be treated in the same manner as real observations. We then apply the emission measure and time lag analyses to this simulated data. \autoref{modeling} provides a detailed description of both our forward modeling pipeline and the nanoflare heating model. In \autoref{results}, we show the predicted intensities for each heating model and AIA channel (\autoref{intensities}), the resulting emission measure slopes (\autoref{em_slopes}) and the time lags (\autoref{timelags}). \autoref{discussion} provides some discussion of our results and \autoref{conclusions} includes a summary and concluding remarks.

This paper is the first in a series concerned with constraining nanoflare heating properties through forward modeled observables and serves to describe our forward modeling pipeline and lay out the results of our nanoflare simulations. In \citet[\citetalias{barnes_understanding_2019-1} hereafter]{barnes_understanding_2019-1}, we use machine learning to make detailed comparisons to AIA observations of \AR{} NOAA 1158. We train a random forest classifier using the predicted emission measure slopes and time lags presented here over the entire heating frequency parameter space in order to classify the heating frequency in each pixel of the observed \AR{}. In contrast to past studies which have relied on a single diagnostic, this approach allows us to simultaneously account for an arbitrarily large number of observables in deciding which model fits the data ``best.'' The ability to quantitatively compare models with large quantities of data is crucial for progress in the current era where the amount of solar coronal data is orders of magnitude larger than in the past.  Combined, these two papers demonstrate a novel method for using real and simulated observations to systematically predict heating properties in \AR{} cores.
 %
%%%%%%%%%%%%%%%%%%%%%%%%%%%%%%%%%%%%%%%%%%%%%%%%%%%%%%%%%%%%%%%%%%%%%%%%%%%%%%%
%                                   Methods                                   %
%%%%%%%%%%%%%%%%%%%%%%%%%%%%%%%%%%%%%%%%%%%%%%%%%%%%%%%%%%%%%%%%%%%%%%%%%%%%%%%
\section{Modeling}\label{modeling}

In order to understand how signatures of the heating frequency are manifested in the emission measure slope and time lag, we predict the emission over the entire \AR{} as observed by SDO/AIA for a range of nanoflare heating frequencies. To do this, we have constructed an advanced forward modeling pipeline through a combination of magnetic field extrapolations, field-aligned hydrodynamic simulations, and atomic data\footnote{Our forward modeling pipeline, called synthesizAR, is modular and flexible and written entirely in Python. The complete source code, along with installation instructions and documentation, are available here: \href{https://github.com/wtbarnes/synthesizAR}{github.com/wtbarnes/synthesizAR}}. In the following section, we discuss each step of our pipeline in detail.

%spell-checker: disable
%spell-checker: enable

%%%%%%%%%%%%%%%%%%%%%%%%%%%%%%%%%%%%%% Field Extrapolation %%%%%%%%%%%%%%%%%%%%
\subsection{Magnetic Field Extrapolation}\label{field}

%spell-checker: disable

\begin{figure*}
	\centering
   \includegraphics[width=\columnwidth]{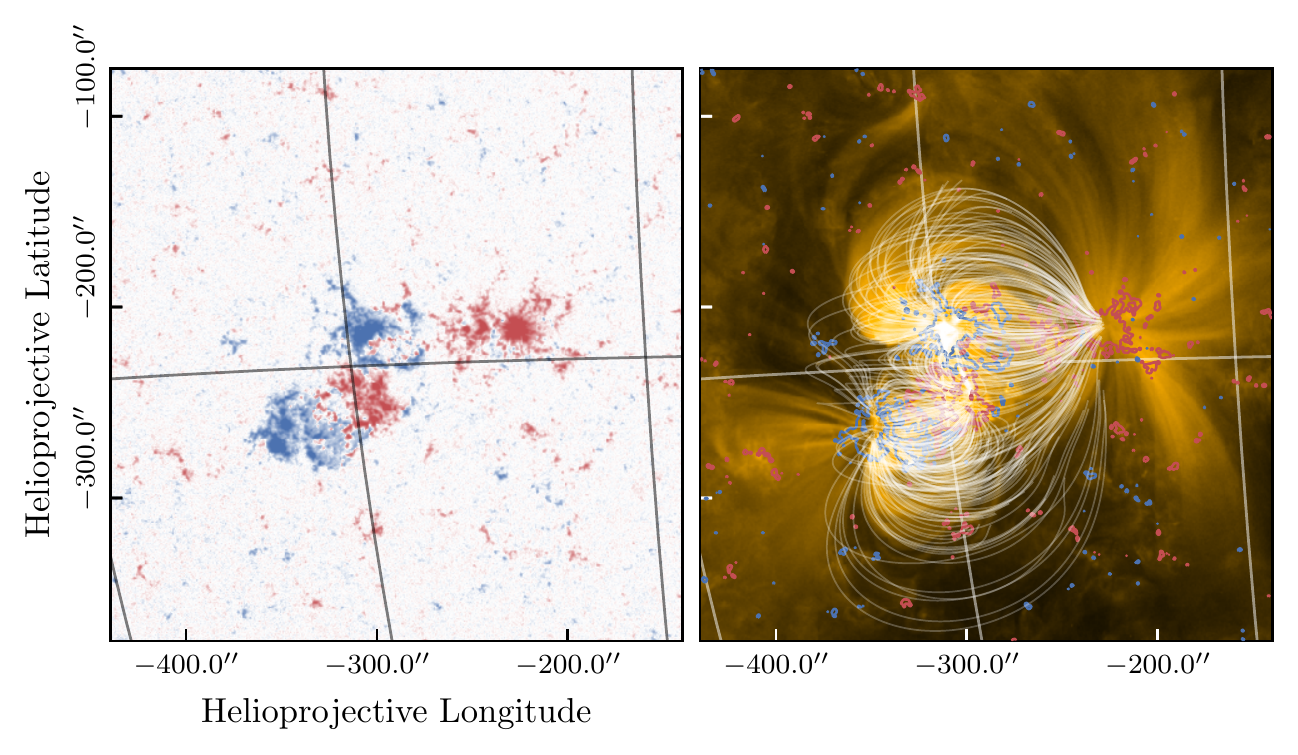}
	\caption{Active region NOAA 1158 on 12 February 2011 15:32:42 UTC as observed by HMI (left) and the 171 \AA{} channel of AIA (right). The gridlines show the heliographic longitude and latitude. The left panel shows the LOS magnetogram and the colorbar range is $\pm750$ G on a symmetrical log scale. In the right panel, 500 out of the total 5000 field lines are overlaid in white and the red and blue contours show the HMI LOS magnetogram at the $+5\%$ (red) and $-5\%$ (blue) levels.}
	\label{fig:magnetogram}
\end{figure*}
%spell-checker:enable

We choose \AR{} NOAA 1158, as observed by the Helioseismic Magnetic Imager \citep[HMI,][]{scherrer_helioseismic_2012} on 12 February 2011 15:32:42 UTC, from the list of active regions studied by \citet{warren_systematic_2012}. The line-of-sight (LOS) magnetogram is shown in the left panel of \autoref{fig:magnetogram}. We model the geometry of \AR{} NOAA 1158 by computing the three-dimensional magnetic field using the oblique potential field extrapolation method of \citet{schmidt_observable_1964} as outlined in \citet[Section 3]{sakurai_greens_1982}. The extrapolation technique of \citeauthor{schmidt_observable_1964} is well-suited for our purposes due to its simplicity and efficiency though we note it is only applicable on the scale of an \AR{}. We include the oblique correction to account for the fact that the \AR{} is off of disk-center.

The HMI LOS magnetogram provides the lower boundary condition of the vector magnetic field (i.e. $B_z(x,y,z=0)$) for our field extrapolation. We crop the magnetogram to an area of 300\arcsec-by-300\arcsec centered on $(-288.26\arcsec,-223.21\arcsec)$ and resample the image to 100-by-100 pixels to reduce the computational cost of the field extrapolation. Additionally, we define our extrapolated field to have a dimension of 100 pixels and spatial extent of $0.3R_{\sun}$ in the $z-$direction such that each component of our extrapolated vector magnetic field $\vec{B}$ has dimensions $(100,100,100)$.

%spell-checker: disable

\begin{figure}
	\centering
   \includegraphics[width=0.5\columnwidth]{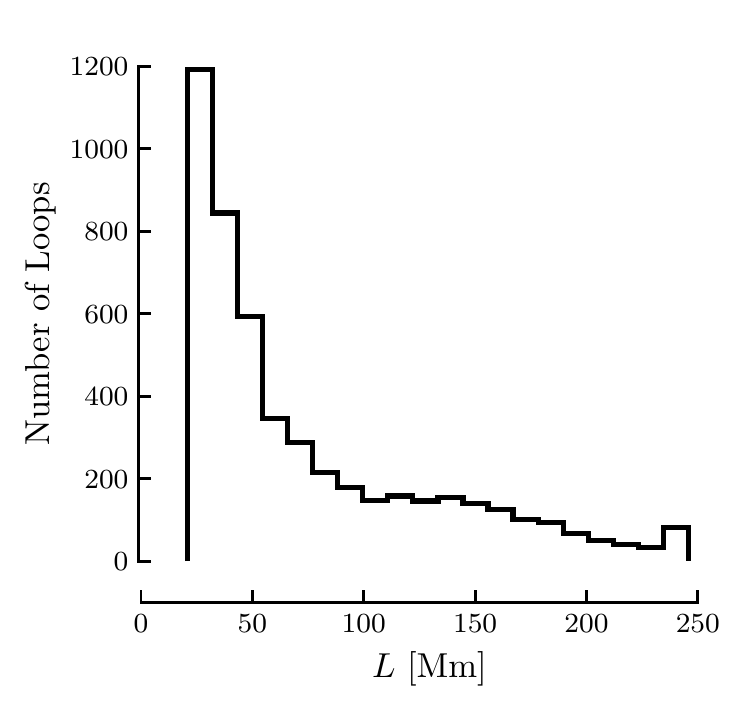}
	\caption{Distribution of footpoint-to-footpoint lengths (in Mm) of the 5000 field lines traced from the field extrapolation computed from the magnetogram of NOAA 1158.}
	\label{fig:loops}
\end{figure}
%spell-checker:enable

After computing the three-dimensional vector field from the observed magnetogram, we trace $5\times10^3$ field lines through the extrapolated volume using the streamline tracing functionality in the yt software package \citep{turk_yt_2011}. \added{We choose $5\times10^3$ lines in order to balance computational cost with the need to make the resulting emission approximately volume filling. We place the seed points for the field line tracing at the lower boundary ($z=0$) of the extrapolated vector field in areas of strong, positive polarity in $B_z$.} Furthermore, we keep only closed field lines in the range $20<L<300$ Mm, where $L$ is the full length of the field line. The right panel of \autoref{fig:magnetogram} shows a subset of the traced field lines overlaid on the observed AIA 171 \AA{} image of NOAA 1158. Contours from the observed HMI LOS magnetogram are shown in red (positive polarity) and blue (negative polarity). A qualitative comparison between the extrapolated field lines and the loops visible in the AIA 171 \AA{} image reveals that the field extrapolation and line tracing adequately capture the three-dimensional geometry of the \AR{}. \autoref{fig:loops} shows the distribution of footpoint-to-footpoint lengths for all of the traced field lines.

%%%%%%%%%%%%%%%%%%%%%%%%%%%%%%%%%%%%% Loop Hydrodynamics %%%%%%%%%%%%%%%%%%%%%%
\subsection{Hydrodynamic Modeling}\label{loops}

Due to the low-$\beta$ nature of the corona, we can treat each field line traced from the field extrapolation as a thermally-isolated strand. We use the Enthalpy-based Thermal Evolution of Loops model \citep[EBTEL,][]{klimchuk_highly_2008,cargill_enthalpy-based_2012,cargill_enthalpy-based_2012-1}, specifically the two-fluid version of EBTEL \citep{barnes_inference_2016}, to model the thermodynamic response of each strand. The two-fluid EBTEL code solves the time-dependent, two-fluid hydrodynamic equations spatially-integrated over the corona for the electron pressure and temperature, ion pressure and temperature, and density. The two-fluid EBTEL model accounts for radiative losses in both the transition region and corona, thermal conduction (including flux limiting), and binary Coulomb collisions between electrons and ions. The time-dependent heating input is configurable and can be deposited in the electrons and/or ions. A detailed description of the model and a complete derivation of the two-fluid EBTEL equations can be found in Appendix B of \citet{barnes_inference_2016}.

For each of the $5\times10^3$ strands, we run a separate instance of the two-fluid EBTEL code for $3\times10^4$ s of simulation time to model the time-dependent, spatially-averaged coronal temperature and density. For each simulation, the loop length is determined from the field extrapolation. We include flux limiting in the heat flux calculation and use a flux limiter constant of 1 \citep[see Equations 21 and 22 of][]{klimchuk_highly_2008}. Additionally, we choose to deposit all of the energy into the electrons \added{though we note that preferentially energizing one species over another will not significantly impact the cooling behavior of the loop as the two species will have had sufficient time to equilibrate \citep{barnes_inference_2016,barnes_inference_2016-1}}. To map the results back to the extrapolated field lines, we assign a single temperature and density to every point along the strand at each time step. Though EBTEL only computes spatially-averaged quantities in the corona, its efficiency allows us to calculate time-dependent solutions for many thousands of strands in a few minutes.

%%%%%%%%%%%%%%%%%%%%%%%%%%%%%%%%%%%%% Heating %%%%%%%%%%%%%%%%%%%%%%%%%%%%%%%%%
\subsection{Heating Model}\label{heating}

We parameterize the heating input in terms of discrete heating pulses on a single strand with triangular profiles of duration $\tau_{\textup{event}}=200$ s. For each event $i$, there are two parameters: the peak heating rate $q_i$ and the waiting time prior to the event $\twait[,i]$. We define the waiting time such that $\twait[,i]$ is the amount of time between when event $i-1$ ends and event $i$ begins. Following the approach of \citet{cargill_active_2014}, we relate the waiting time and the event energy such that $\twait[,i]\propto q_i$. The physical motivation for this scaling is as follows. In the nanoflare model of \citet{parker_nanoflares_1988}, random convective motions continually stress the magnetic field rooted in the photosphere, leading to the buildup and eventual release of energy. If the field is stressed for a long amount of time without relaxation, large discontinuities will have time to develop in the field, leading to a dramatic release of energy. Conversely, if the field relaxes quickly, there is not enough time for the field to become sufficiently stressed and the resulting energy release will be relatively small. 

In this work we explore three different heating scenarios: low-, intermediate-, and high-frequency nanoflares. We define the heating frequency in terms of the ratio between the fundamental cooling timescale due to thermal conduction and radiation, $\tau_{\textup{cool}}$, and the average waiting time of all events on a given strand, $\langle \twait\rangle$,

\begin{equation}\label{eq:heating_types}
    \varepsilon = \frac{\langle \twait\rangle}{\tau_{\textup{cool}}}
    \begin{cases} 
        < 1, &  \text{high frequency},\\
        \sim1, & \text{intermediate frequency}, \\
        > 1, & \text{low frequency}.
     \end{cases}
\end{equation}

\begin{figure}
	\centering
   \includegraphics[width=0.5\columnwidth]{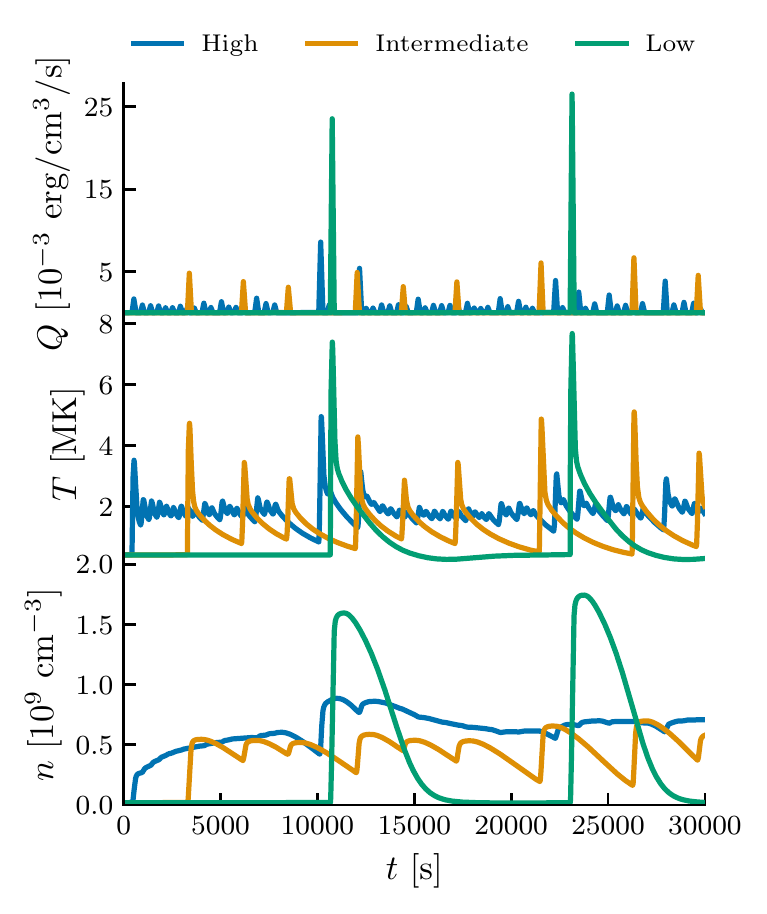}
	\caption{Heating rate (top), electron temperature (middle), and density (bottom) as a function of time for the three heating scenarios for a single strand. The colors denote the heating frequency as defined in the legend. The strand has a half length of $L/2\approx40$ Mm and a mean field strength of $\bar{B}\approx30$ G.}
	\label{fig:hydro-profiles}
\end{figure}

We choose to parameterize the heating in terms of the cooling time rather than an absolute waiting time as $\tau_{\textup{cool}}\sim L$ \citep[see appendix of][]{cargill_active_2014}. While a waiting time of 2000 s might correspond to low-frequency heating for a 20 Mm strand, it would correspond to high-frequency heating in the case of a 150 Mm strand. By parameterizing the heating in this way, we ensure that all strands in the \AR{} are heated at the same frequency relative to their cooling time. \autoref{fig:hydro-profiles} shows the heating rate, electron temperature, and density as a function of time, for a single strand, for the three heating scenarios listed above. 

% spell-checker: disable %

% spell-checker: enable %

For a single impulsive event $i$ with a triangular temporal profile of duration $\tau_{\textup{event}}$, the energy density is $E_i=\tau_{\textup{event}}q_i/2$. Summing over all events on all strands that comprise the \AR{} gives the total energy flux injected into the \AR{},
\begin{equation}
    F_{AR} = \frac{\tau_{\textup{event}}}{2}\frac{\sum_l^{N_{\textup{strands}}}\sum_i^{N_l} q_iL_l}{t_\textup{total}}
\end{equation}
where $t_\textup{total}$ is the total simulation time, $N_\textup{strands}$ is the total number of strands comprising the \AR{}, and $N_l=(t_\textup{total} + \langle\twait\rangle)/(\tau + \langle\twait\rangle)$ is the total number of events occurring on each strand over the whole simulation. Note that the number of events per strand is a function of both $\varepsilon$ and $\tau_{\textup{cool}}$.

For each heating frequency, we constrain the total flux into the \AR{} to be $F_{\ast}=10^7$ erg cm$^{-2}$ s$^{-1}$ \citep{withbroe_mass_1977} such that $F_{AR}$ must satisfy the condition,
\begin{equation}\label{eq:energy_constraint}
    \frac{| F_{AR}/N_\textup{strands} - F_{\ast} |}{F_{\ast}} < \delta,
\end{equation}
where $\delta\ll1$. For each strand, we choose $N_l$ events each with energy $E_i$ from a power-law distribution with slope $-2.5$ and fix the upper bound of the distribution to be $\bar{B}_l^2/8\pi$, where $\bar{B}_l$ is the spatially-averaged field strength along the strand $l$ as derived from the field extrapolation. This is the maximum amount of energy made available by the field to heat the strand. We then iteratively adjust the lower bound on the power-law distribution for $E_i$ until we have satisfied \autoref{eq:energy_constraint} within some numerical tolerance. We note that the set of $E_i$ we choose for each strand may not uniquely satisfy \autoref{eq:energy_constraint}.

We use the field strength derived from the potential field extrapolation to constrain the energy input to our hydrodynamic model for each strand. While the derived potential field is already in its lowest energy state and thus has no energy to give up, our goal here is only to understand how the distribution of field strength may be related to the properties of the heating. In this way, we use the potential field as a proxy for the non-potential component of the coronal field, with the understanding that we cannot make any quantitative conclusions regarding the amount of available energy or the stability of the field itself.

\begin{deluxetable}{lcc}
    \tablecaption{All three heating models plus the two single-event control models. In the single-event models, the energy flux is not constrained by \autoref{eq:energy_constraint}.\label{tab:heating}}
    \tablehead{\colhead{Name} & \colhead{$\varepsilon$ (see Eq.\ref{eq:heating_types})} & \colhead{Energy Constrained?}}
    \startdata
    high & 0.1 & yes \\
    intermediate & 1 & yes \\
    low & 5 & yes \\
    cooling & 1 event per strand & no \\
    random & 1 event per strand & no
    \enddata
\end{deluxetable}

In addition to these three multi-event heating models, we also run two single-event control models. In both control models every strand in the \AR{} is heated exactly once by an event with energy $\bar{B}_l^2/8\pi$. In our first control model, the start time of every event is $t=0$ s such that all strands are allowed to cool uninterrupted for $t_\textup{total}=10^4$ s. In the second control model, the start time of the event on each strand is chosen from a uniform distribution over the interval $[0, 3\times10^4]$ s, such that the heating is likely to be out of phase across all strands. In these two models, the energy has not been constrained according to \autoref{eq:energy_constraint} and the total flux into the \AR{} is $(\sum_{l}\bar{B}_l^2L_l)/8\pi t_\textup{total}$. From here on, we will refer to these two models as the ``cooling'' and ``random'' models, respectively. All five heating scenarios are summarized in \autoref{tab:heating}.

%%%%%%%%%%%%%%%%%%%%%%%%%%%%%%%%%%%%%%% Forward Modeling %%%%%%%%%%%%%%%%%%%%%%
\subsection{Forward Modeling}\label{forward}

\subsubsection{Atomic Physics}\label{atomic}

For an optically-thin, high-temperature, low-density plasma, the radiated power per unit volume, or \textit{emissivity}, of a transition $\lambda_{ij}$ of an electron in ion $k$ of element $X$ is given by,
\begin{equation}
    \label{eq:ppuv}
    P(\lambda_{ij}) = \frac{n_H}{n_e}\mathrm{Ab}(X)N_j(X,k)f_{X,k}A_{ji}\Delta E_{ji}n_e,
\end{equation}
where $N_j$ is the fractional energy level population of excited state $j$, $f_{X,k}$ is the fractional population of ion $k$, $\mathrm{Ab}(X)$ is the abundance of element $X$ relative to hydrogen, $n_H/n_e\approx0.83$ is the ratio of hydrogen and electron number densities, $A_{ji}$ is the Einstein coefficient, and $\Delta E_{ji}=hc/\lambda_{ij}$ is the energy of the emitted photon \citep[see][]{mason_spectroscopic_1994,del_zanna_solar_2018}. To compute \autoref{eq:ppuv}, we use version 8.0.6 of the CHIANTI atomic database \citep{dere_chianti_1997,young_chianti_2016}. We use the abundances of \citet{feldman_potential_1992} as provided by CHIANTI. For each atomic transition, $A_{ji}$ and $\lambda_{ji}$ can be looked up in the database. To find $N_j$, we solve the level-balance equations for ion $k$, including the relevant excitation and de-excitation processes as provided by CHIANTI \citep[see Section 3.3 of][]{del_zanna_solar_2018}.

The ion population fractions, $f_{X,k}$, provided by CHIANTI assume ionization equilibrium (i.e. the ionization and recombination rates are always in balance). However, in the rarefied solar corona, where the plasma is likely heated impulsively, it is not guaranteed that the ionization timescale is less than the heating timescale such that the ionization state may not be representative of the electron temperature \citep{bradshaw_explosive_2006,reale_nonequilibrium_2008,bradshaw_numerical_2009}. To properly account for this effect, we compute $f_{X,k}$ by solving the time-dependent ion population equations for each element using the ionization and recombination rates provided by CHIANTI. The details of this calculation are provided in \autoref{nei}.

\subsubsection{Instrument Effects}\label{instrument}

% spell-checker: disable %
\begin{deluxetable}{ccc}
\tablecaption{Elements included in the calculation of \autoref{eq:intensity}. For each element, we include all ions for which CHIANTI provides sufficient data for computing the emissivity.\label{tab:elements}}
\tablehead{\colhead{Element} & \colhead{Number of Ions} & \colhead{Number of Transitions}}
\startdata
O & 8 & 11892 \\
Mg & 11 & 31965 \\
Si & 13 & 30047 \\
S & 16 & 33091 \\
Ca & 17 & 42823 \\
Fe & 25 & 553541 \\
Ni & 19 & 83517
\enddata
\end{deluxetable}
% spell-checker: enable %

We combine \autoref{eq:ppuv} with the wavelength response function of the instrument to model the intensity as it would be observed by AIA,
\begin{equation}\label{eq:intensity}
    I_c = \frac{1}{4\pi}\sum_{\{ij\}}\int_{\text{LOS}}\mathrm{d}hP(\lambda_{ij})R_c(\lambda_{ij})
\end{equation}
where $I_c$ is the intensity for a given pixel in channel $c$, $P(\lambda_{ij})$ is the emissivity as given by \autoref{eq:ppuv}, $R_c$ is the wavelength response function of the instrument for channel $c$ \citep[see][]{boerner_initial_2012}, $\{ij\}$ is the set of all atomic transitions listed in \autoref{tab:elements}, and the integration is along the LOS. Note that when computing the intensity in each channel of AIA, we do not rely on the temperature response functions computed by SolarSoft \citep[SSW,][]{freeland_data_1998} and instead use the wavelength response functions directly. This is because the response functions returned by \texttt{aia\_get\_response.pro} assume both ionization equilibrium and constant pressure. \autoref{effective_response_functions} provides further details on our motivation for recomputing the temperature response functions.

We compute the emissivity according to \autoref{eq:ppuv} for all of the transitions in \autoref{tab:elements} using the temperatures and densities from from our hydrodynamic models for all $5\times10^3$ strands. We then compute the LOS integral in \autoref{eq:intensity} by first converting the coordinates of each strand to a helioprojective (HPC) coordinate frame \citep[see][]{thompson_coordinate_2006} using the coordinate transformation functionality in Astropy \citep{the_astropy_collaboration_astropy_2018} combined with the solar coordinate frames provided by SunPy \citep{sunpy_community_sunpypython_2015}. This enables us to easily project our simulated \AR{} along any arbitrary LOS simply by changing the location of the observer that defines the HPC frame. Here, our HPC frame is defined by an observer at the position of the SDO spacecraft on 12 February 2011 15:32:42 UTC (i.e. the time of the HMI observation of NOAA 1158 shown in \autoref{fig:magnetogram}).

Next, we use these transformed coordinates to compute a weighted two-dimensional histogram, using the integrand of \autoref{eq:intensity} at each coordinate as the weights. We construct the histogram such that the bin widths are consistent with the spatial resolution of the instrument. For AIA, a single bin, representing a single pixel, has a width of 0.6\arcsec-per-pixel. Finally, we  apply a gaussian filter to the resulting histogram to emulate the point spread function of the instrument. We do this for each time step, using a cadence of 10 s, and for each channel. For every heating scenario, this produces approximately $6(3\times10^4)/10\approx2\times10^4$ separate images.
 %
%%%%%%%%%%%%%%%%%%%%%%%%%%%%%%%%%%%%%%%%%%%%%%%%%%%%%%%%%%%%%%%%%%%%%%%%%%%%%%%
%                                   Results                                   %
%%%%%%%%%%%%%%%%%%%%%%%%%%%%%%%%%%%%%%%%%%%%%%%%%%%%%%%%%%%%%%%%%%%%%%%%%%%%%%%
\section{Results}\label{results}

% spell-checker: disable %
% spell-checker: enable %

We forward model time-dependent AIA intensities using the method outlined in \autoref{forward} for the heating scenarios discussed in \autoref{heating}. We discuss the predicted intensities in \autoref{intensities} for all six EUV channels of AIA and all five heating models. In \autoref{em_distributions} and \autoref{timelags}, we show the results of the emission measure and time lag analyses, respectively, as applied to our simulated data. In \citetalias{barnes_understanding_2019-1}, we use these simulated observables to train a machine learning classification model to understand with which heating scenario the real data are most consistent.

%%%%%%%%%%%%%%%%%%%%%%%%%%%%%%%%%%%%%%% Intensities %%%%%%%%%%%%%%%%%%%%%%%%%%%
\subsection{Intensities}\label{intensities}

We compute the intensities for the 94, 131, 171, 193, 211, and 335 \AA{} channels of SDO/AIA using the procedure described in \autoref{forward}. We compute the intensity in each pixel of the model \AR{} over a total simulation period of $3\times10^4\,\mathrm{s}\,\approx8.3$ hours with the exception of the cooling case which is only run for $10^4$ s. For the high-, intermediate-, low-frequency and ``random'' models, we discard the first and last $5\times10^3$ s of evolution to avoid any transient effects in the strand evolution associated with the initial conditions and the constraints placed on the energy, respectively. We complete this procedure for each of the five heating scenarios in \autoref{tab:heating}.

% spell-checker: disable %

\begin{figure*}
	\centering
   \includegraphics[width=\columnwidth]{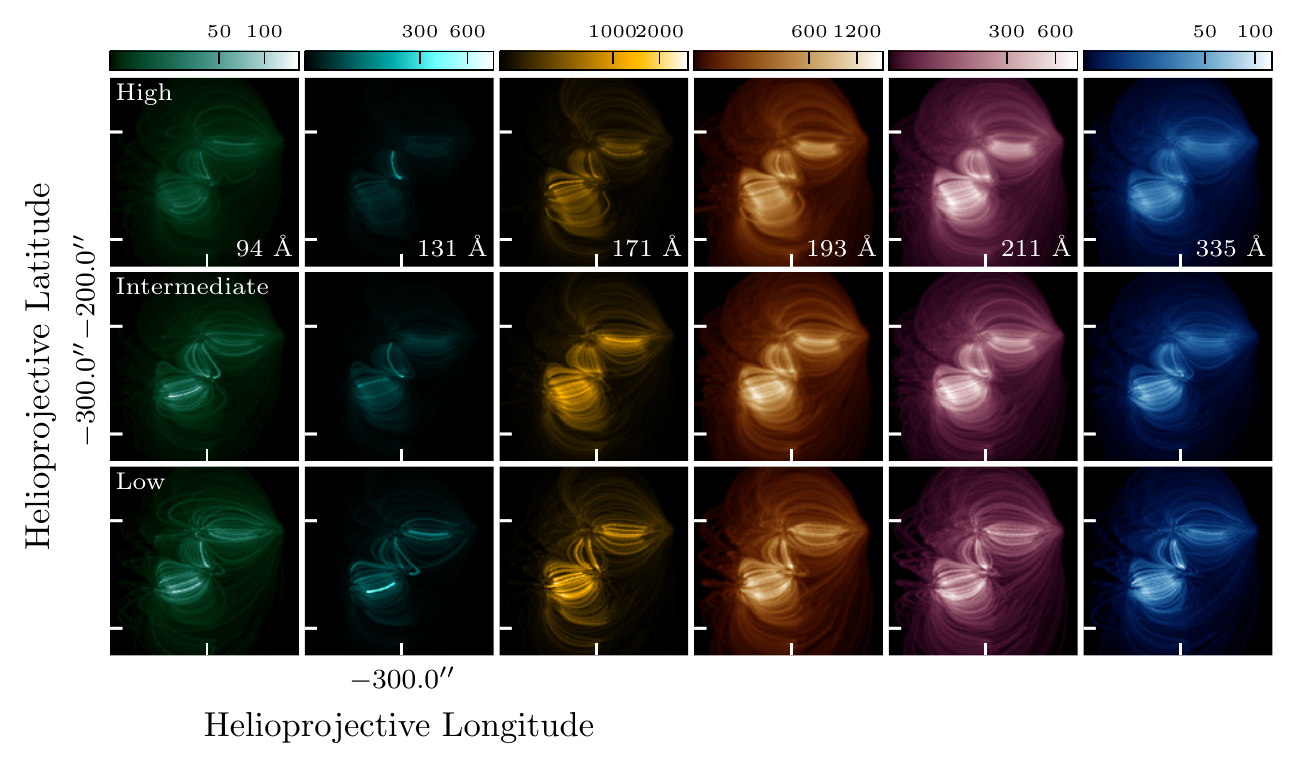}
	\caption{Snapshots of intensity, in DN pixel$^{-1}$ s$^{-1}$, across the whole \AR{} at $t=15\times10^3$ s. The rows correspond to the three different heating frequencies and the columns are the six EUV channels of AIA. In each column, the colorbar is on a square-root scale and is normalized between zero and the maximum intensity in the low-frequency case. The color tables are the standard AIA color tables as implemented in SunPy \citep{sunpy_community_sunpypython_2015}.}
	\label{fig:intensity-map}
\end{figure*}
% spell-checker: enable %

\autoref{fig:intensity-map} shows a snapshot of the intensity map at $t=15\times10^3$ s for each channel and for the high-, intermediate-, and low-frequency nanoflare heating cases. The rows correspond to the different heating scenarios while the columns show the six AIA EUV channels. In each column, the intensities are normalized to the maximum intensity in the low-frequency case and are on a square-root scale. In general, we find that in the high-frequency intensity maps, individual loops are difficult to distinguish while in the low-frequency case individual loops appear bright relative to the surrounding emission. This distinguishability or ``fuzziness'' can be measured quantitatively as $\sigma_{I}/\bar{I}$, where $\sigma_{I}$ is the standard deviation taken over all pixels and $\bar{I}$ is the mean intensity \citep[Equation 11]{guarrasi_coronal_2010}. A larger value of $\sigma_{I}/\bar{I}$ indicates a greater degree of contrast and vice versa. $\sigma_{I}/\bar{I}$ for each channel and heating frequency is shown in \autoref{tab:fuzzy}. With the exception of 131 \AA{}, for every channel, the high-frequency case is the most ``fuzzy''. The low-frequency case shows the most contrast in each channel except 94 \AA{} though the margin between the low and intermediate cases is quite small in some cases.

% spell-checker: disable %
\begin{deluxetable}{cccc}
\tablecaption{$\sigma_I/\bar{I}$ as defined by Equation 11 of \citet{guarrasi_coronal_2010} computed on a single image at $t=15\times10^3$ s for each channel and heating frequency. A larger value denotes a greater degree of contrast.\label{tab:fuzzy}}
\tablehead{\colhead{Channel [\AA]} & \colhead{High} & \colhead{Intermediate} & \colhead{Low}}
\startdata
94 & 3.06 & 4.63 & 4.19 \\
131 & 5.56 & 3.61 & 6.13 \\
171 & 2.79 & 2.81 & 3.25 \\
193 & 2.69 & 2.80 & 2.79 \\
211 & 2.73 & 2.83 & 2.84 \\
335 & 2.63 & 3.08 & 3.21
\enddata
\end{deluxetable}
% spell-checker: enable %

Looking at the first two columns of \autoref{fig:intensity-map}, we see that the intensity in the 94 and 131 \AA{} channels increases as the heating frequency decreases. Both channels are double peaked and have ``hot'' ($\approx7$ MK for 94 \AA{}, $\approx12$ MK for 131 \AA{}) and ``warm'' ($\approx1$ MK for 94 \AA{}, $\approx0.5$ MK for 131 \AA{}) components. In the case of high-frequency heating, less energy is available per event such that few strands are heated to $>4$ MK. There is little emission in the 131 \AA{} channel as strands are not often permitted to cool to $\leq0.5$ MK either. However, in the low- and intermediate-frequency cases, we see several individual bright loops in both the 94 and 131 \AA{} channels as the heating rate is sufficient to produce ``hot'' (i.e. 8-10 MK) loops. We see only a few of these loops as the lifetime of this hot plasma is short due to the efficiency of thermal conduction. In contrast, the faint, diffuse component of the 94 \AA{} emission that is present in all three cases is due to the contribution of the ``warm'' component. 

Additionally, we find that the 171 \AA{} channel is dimmer for high-frequency heating as the peak sensitivity of this channel is $<1$ MK and in the case of high-frequency heating, strands are rarely allowed to cool below $1$ MK. In contrast, we note that the overall intensity in the 193, 211, and 335 \AA{} channels is relatively constant over heating frequency as compared to the three previous channels though individual loops do become more visible with decreasing heating frequency. This relative insensitivity is because the temperature response functions of these three channels all peak in between 1.5 MK and 2.5 MK. In the case of high-frequency heating, strands are being sustained near these temperatures while in the low-frequency case, strands are cooling through this temperature range. This is illustrated for a single strand in \autoref{fig:hydro-profiles}.

While there are clear differences in the AIA intensities between all three heating frequencies, quantifying these differences is difficult due in part to the multidimensional nature of the intensity data. To better understand how observational signatures differ as a function of heating frequency, we need to find a reduced representation of our data set that retains signatures of the underlying energy deposition. To this end, we compute two common observables: the emission measure slope (\autoref{em_distributions}) and the time lag (\autoref{timelags}).

%%%%%%%%%%%%%%%%%%%%%%%%%%%%%%%%%%%%%%% Emission Measure Distributions %%%%%%%%%%%%%%%%
\subsection{Emission Measure Distributions}\label{em_distributions}

As discussed in \autoref{introduction}, the emission measure slope is a useful quantity for understanding how frequently strands are re-energized. We compute emission measure distributions from our forward-modeled intensities using the regularized inversion method of \citet{hannah_differential_2012}. This method was designed to work with the narrowband coverage provided by AIA and so is well-suited to our needs. We choose our temperature bins such that the leftmost edge is at $10^{5.5}$ K and the rightmost edge at $10^{7.2}$ K with bin widths of $\Delta\log T=0.1$. Rather than computing \dem{} at each time step, we compute the time-averaged intensity in each pixel of each channel and compute \dem{} only once. We compute the uncertainties on the intensities in each channel using the \texttt{aia\_bp\_estimate\_error.pro} procedure in SSW which incorporates uncertainties due to shot noise, read noise, dark subtraction, quantization, photometric calibration, and onboard compression.

% spell-checker: disable %

\begin{figure*}
	\centering
   \includegraphics[width=\columnwidth]{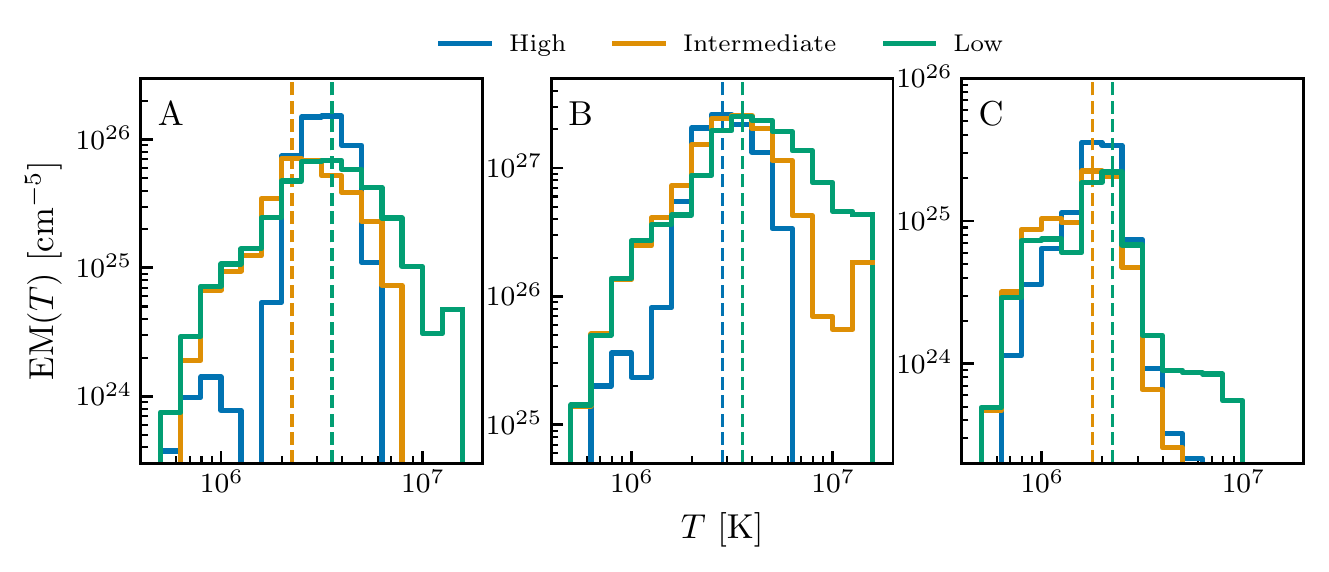}
	\caption{\dem{} at three different locations in the \AR{} for all three heating frequencies. The dashed vertical lines denote the value of $\Tpeak$. Each panel corresponds to a different pixel in the \AR{} as denoted by the label in the top left corner. The locations of these three pixels are marked with their corresponding labels in \autoref{fig:em-slope-maps}.}
	\label{fig:em-dist}
\end{figure*}
% spell-checker: enable %

\added{\autoref{fig:em-dist} shows the emission measure distribution, \dem{}, for the case of high-, intermediate-, and low-frequency nanoflares. The three panels correspond to the three locations marked in \autoref{fig:em-slope-maps}. The dashed lines in each panel denote the location of $\Tpeak$, where $\Tpeak=\argmax_T\,\mathrm{EM}(T)$ is the temperature at which the emission measure distribution is maximum. At each sample location, the \dem{} becomes increasingly narrow with increasing heating frequency. We find that $\Tpeak$ is between $\approx2$ MK and 4 MK in all cases and is lowest at point C near the periphery of the \AR{} where the loops are the longest. We note that in all cases, $\Tpeak$ is significantly below 4 MK, the value measured by \citet{warren_systematic_2012} from spectroscopic observations of this same \AR{}. However, we note that at point B, the location closest to that at which \citeauthor{warren_systematic_2012} computed their \dem{} distributions, we find the highest $\Tpeak$, between $\approx3$ MK and 4 MK.}

\subsubsection{Emission Measure Slopes}\label{em_slopes}

After computing \dem{} in each pixel using the regularized inversion procedure, we fit a first-order polynomial to the log-transformed emission measure and temperature bin centers, $\log_{10}\mathrm{EM}\sim a\log_{10}T$, to calculate the emission measure slope, $a$. In \citetalias{barnes_understanding_2019-1}, we compare our modeled emission measure slopes to those derived from real AIA observations of NOAA 1158 using this same method.

% spell-checker: disable %

\begin{figure*}
	\centering
   \includegraphics[width=\columnwidth]{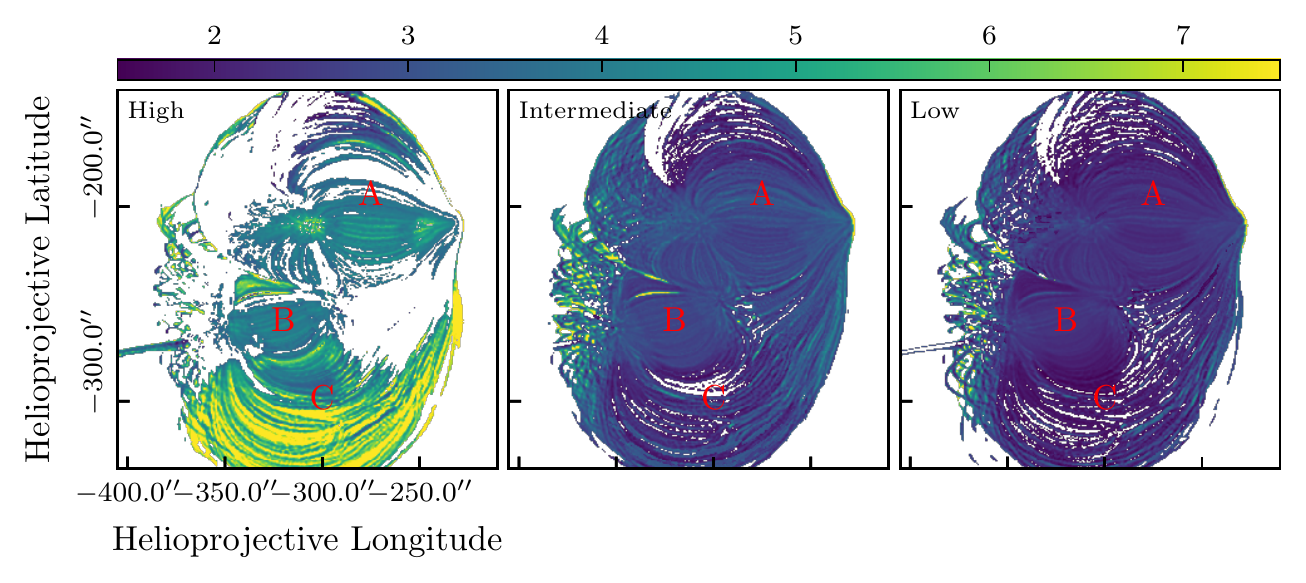}
	\caption{Maps of the emission measure slope, $a$, in each pixel of the \AR{} for the high- (left), intermediate- (center), and low-frequency (right) cases. The \dem{} is computed using time-averaged intensities from the six AIA EUV channels using the method of \citet{hannah_differential_2012}. The \dem{} in each pixel is then fit to $T^a$ over the temperature range $8\times10^{5}\,\textup{K}\le T< \Tpeak$. Any pixels with $r^2<0.75$ are masked and colored white. The red labels in each panel denote the three locations at which the \dem{} distributions in \autoref{fig:em-dist} were calculated.}
	\label{fig:em-slope-maps}
\end{figure*}
% spell-checker: enable %

\autoref{fig:em-slope-maps} shows the resulting emission measure slope, $a$, in each pixel of our forward-modeled \AR{} for the high-, intermediate-, and low-frequency cases. We fit \dem{} over bins in the temperature range $8\times10^{5}\,\textup{K}\le T< \Tpeak$. To assess the ``goodness-of-fit'' we use $r^2$, the correlation coefficient for the first-order polynomial fit, and mask pixels with $r^2<0.75$. Looking at the three panels in \autoref{fig:em-slope-maps}, we find that overall, $a$ tends to decrease with decreasing frequency, consistent with previous modeling work (see \autoref{introduction}). The low-frequency map (right panel) shows many values close to 2.

As the heating frequency increases, the slopes become larger, indicating an increasingly isothermal emission measure distribution. The intermediate-frequency map (center panel) shows predominantly higher slopes, with most pixels in the range $2\lesssim a \lesssim 3.5$ while the high-frequency map (left panel) shows much steeper slopes, with many $a\ge3.5$, and a much broader range of slopes, $3\lesssim a \lesssim 8$. Note that in the high-frequency case, the slope varies considerably across the \AR{} while the distribution of $a$ appears more spatially uniform in the intermediate- and low-frequency cases.

Below $\Tpeak$, \citet{cargill_implications_1994} noted that the \dem{} could be described by $\mathrm{EM}(T)\sim n^2\tau_{rad}$, where $\tau_{rad}\sim T^{1-\alpha}n^{-1}$ is the radiative cooling time \added{and $\alpha$ determines the temperature dependence of the radiative loss function}. \deleted{Additionally, \citet{bradshaw_cooling_2010} found that $T\sim n^{\ell}$, with $\ell\approx1$ for long loops and $\ell\approx2$ for short loops.} \added{Assuming $T\propto n^2$ \citep{serio_dynamics_1991,jakimiec_dynamics_1992} and approximating the temperature dependence of the radiative losses as $\alpha=-1/2$ \citep{cargill_implications_1994,cargill_cooling_1995} gives $a\approx2$.} We find that emission measure slopes produced by low-frequency nanoflares as shown in the right panel of \autoref{fig:em-slope-maps} are \added{approximately} consistent with analytical results for single nanoflares \added{though many of the low-frequency slopes are $>2$}. \added{This is likely due to the omission of enthalpy-driven cooling in the above scaling relation. \citet{bradshaw_cooling_2010} found $T\sim n^{\ell}$, with $\ell\approx1$ for long loops, where enthalpy-driven cooling is likely to dominate over radiative losses, and $\ell\approx2$ for shorter loops, where radiation remains the dominant cooling mechanism. Thus, the distribution of $a$ will depend on the distribution of loop lengths and smaller values of $\ell$ will lead to larger emission measure slopes (e.g. $a=2.5$ for $\ell=1$).}

\subsubsection{Histograms}\label{em_histograms}

% spell-checker: disable %

\begin{figure}
	\centering
   \includegraphics[width=0.5\columnwidth]{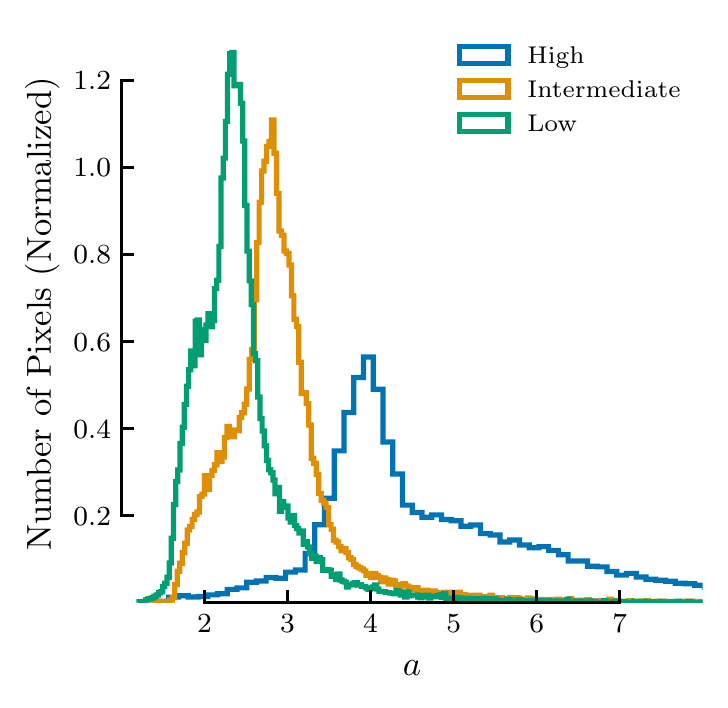}
	\caption{Distribution of emission measure slopes, $a$, for every pixel in the simulated \AR{} for the high-, intermediate-, and low-frequency heating scenarios as shown in \autoref{fig:em-slope-maps}. The histogram bins are determined using the Freedman Diaconis estimator \citep{freedman_histogram_1981} as implemented in the Numpy package for array computation in Python \citep{oliphant_guide_2006} and each histogram is normalized such that the area under the histogram is equal to 1.}
	\label{fig:em-slope-histograms}
\end{figure}
% spell-checker: enable %

\autoref{fig:em-slope-histograms} shows histograms of the emission measure slopes for the high-, intermediate-, and low-frequency cases. We find that the low-frequency distribution peaks at $a\approx2.3$, inside the range expected from analytical results as noted above. The intermediate- and high-frequency distributions peak at successively higher values, $\approx2.8$ and $\approx4.0$, respectively. While the low- and intermediate-frequency distributions are more narrowly distributed around their peak values, the distribution of slopes in the high-frequency case is relatively broad and has a positive skew towards steeper slopes. 

Looking at the distribution of slopes across the entire \AR{} in \autoref{fig:em-slope-histograms}, we find that when the strands are heated infrequently such that each strand is allowed to cool fully prior to the next event, the distribution of slopes ``saturates'' in the range expected for single nanoflares. However, as the \added{heating frequency increases and the }strands are reheated more often, the value of the slope becomes unsaturated and is subject to a \added{wide} range of \deleted{infrequent} cooling times due to the dependence of each waiting time on the power-law heating rate (see \autoref{heating}). These results are consistent with \citet{cargill_active_2014} who computed $\mathrm{EM}(T)=n^2L$ for a single strand for a range of heating frequencies and found $a$ converged to $\approx2$ for low frequency nanoflares and increased slowly with increasing heating frequency.

Our modeled emission measure slopes show that, even when accounting for the LOS integration, atomic physics, and information lost in the \dem{} inversion, signatures of the heating frequency still persist in the emission measure slope. However, while this quantity retains information about the frequency of energy deposition, drawing conclusions about the heating based solely on the observed emission measure slope, particularly for a small number of pixels may be misleading. As shown here and in \citet{del_zanna_evolution_2015}, the slope may vary significantly across a given \AR{}. Additionally, calculating \dem{} from observations is non-trivial due to several factors, including the mathematical difficulties of the ill-posed inversion \citep{craig_fundamental_1976,judge_failure_1995,judge_fundamental_1997}, uncertainties in the atomic data \citep{guennou_can_2013}, and insufficient constraints from spectroscopic observations \citep[e.g.][]{landi_isothermality_2010,winebarger_defining_2012}, among others.

%%%%%%%%%%%%%%%%%%%%%%%%%%%%%%%%%%%%%%% Time lags %%%%%%%%%%%%%%%%%%%%%%%%%%%%
\subsection{Time Lags}\label{timelags}

Next, we apply the time lag method of \citet{viall_evidence_2012} to our synthetic intensities for all of the heating scenarios discussed in \autoref{heating}. For each pixel in the active region, we compute the cross-correlation (\autoref{eq:cc}) for all pairs of the EUV channels of AIA (15 in total) and find the temporal offset which maximizes the cross-correlation according to \autoref{eq:timelag}. The details of the cross-correlation and time lag calculation are given in \autoref{timelag_details}. \deleted{We consider all possible offsets over the interval $\pm6$ hours.} Using the convention of \citet{viall_evidence_2012}, we order the channel pairs such that the ``hot'' channel is listed first, meaning that a positive time lag corresponds to variability in the hotter channel followed by variability in the cooler channel. In other words, \textit{a positive time lag indicates cooling plasma.} For the 94 \AA{} and 131 \AA{} channels, both of which have a bimodal temperature response function (see \autoref{fig:aia-response}), the order is determined by the component which is most dominant such that 94 \AA{} is ordered first while 131 \AA{} is ordered second. Thus, it is possible for cooling plasma to produce negative time lags in these channel pairs and the ambiguity can be resolved in the context of the time lags in other channel pairs.

\subsubsection{Time Lag Maps}\label{timelag_maps}

% spell-checker: disable %

\begin{figure*}
	\centering
   \includegraphics[width=\columnwidth]{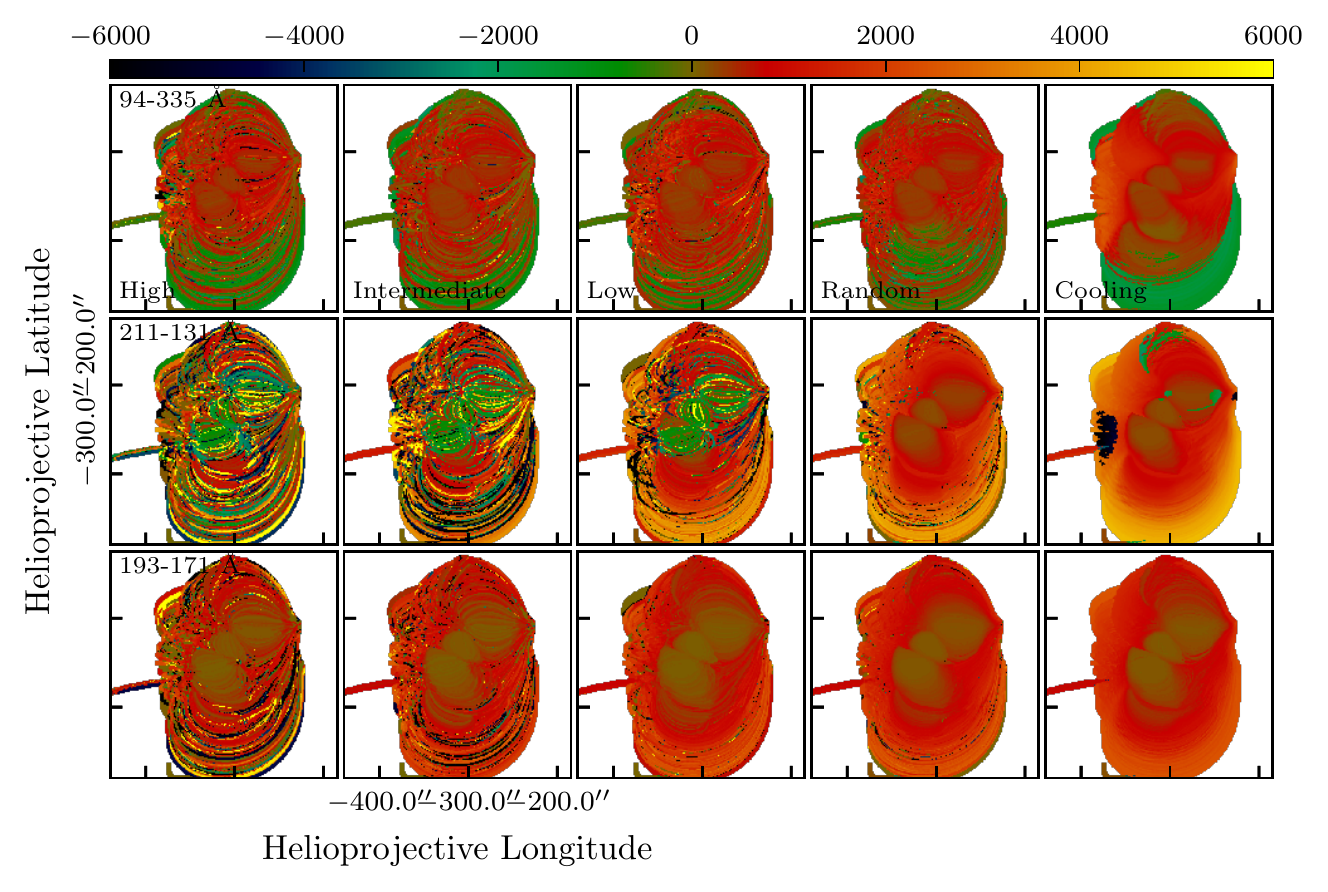}
	\caption{Time lag maps for three different channel pairs for all five of the heating scenarios described in \autoref{tab:heating}. The value of each pixel indicates the temporal offset, in seconds, which maximizes the cross-correlation (\autoref{eq:timelag}). The rows indicate the different channel pairs and the columns indicate the varying heating scenarios. The range of the colorbar is $\pm6000$ s. If $\max{\mathcal{C}_{AB}}<0.1$, the pixel is masked and colored white.}
	\label{fig:timelag-maps}
\end{figure*}
% spell-checker: enable %

\autoref{fig:timelag-maps} shows $\tau_{AB}$ (\autoref{eq:timelag}) in each pixel of our simulated \AR{} for all heating scenarios listed in \autoref{tab:heating} and three selected channel pairs: 94-335 \AA{}, 211-131 \AA{}, and 193-171 \AA{}. Blacks, blues, and greens correspond to negative time lags; reds, oranges, and yellows correspond to positive time lags; and olive green indicates zero time lag. The range of the colorbar is $\pm6000$ s. Note that the heating frequency decreases from left to right across each row. If the correlation in a given pixel is too low ($\max{\mathcal{C}_{AB}}<0.1$), the pixel is masked and colored white.

Looking at the first two rows of \autoref{fig:timelag-maps}, we find that the positive time lags in the 211-131 \AA{} channel pair are significantly longer than those in the 94-335 \AA{} pair. In the temperature range $2.5<T<7.3$ MK (94-335 \AA{}), the dominant cooling mechanism is field-aligned thermal conduction while radiative cooling dominates in the range $0.6<T<2.5$ MK (211-131 \AA{}). Because thermal conduction is far more efficient at high temperatures, the plasma spends less time in the $[T_{335},T_{94}]$ temperature range than in $[T_{131},T_{211}]$. The 193-171 \AA{} time lags for the cooling case fall in the middle as radiative cooling also tends to dominate in this temperature interval ($0.9<T<1.5$ MK), but the separation in temperature space is smaller than the 211-131 \AA{} pair. In all cases, these differences in the magnitude of the positive time lags become more apparent at lower heating frequencies.

In the 94-335 \AA{} pair, we find negative time lags in the longest loops near the edge of the \AR{}, inconsistent with our previous assertion that longer loops lead to longer, positive time lags. These loops are rooted in areas of weaker magnetic field (compared to the center) and thus do not have sufficient energy to evolve significantly into the temperature range of the ``hot'' component of the 94 \AA{} channel (see \autoref{heating}). Thus, cooling from 335 \AA{} to the cooler part of 94 \AA{} dominates the cross-correlation. These negative time lags become more prominent as the heating frequency decreases. \replaced{Our results in the cooling case are consistent with the negative 94-335 \AA{} time lags of similar magnitude observed by \citet{viall_survey_2017} in this same active region and the tendency for longer field lines to exhibit cooler plasma,}{Consistent with the observations by \citet{viall_survey_2017} of this same \AR{}, our simulation results for the ``cooling'' scenario (rightmost column of \autoref{fig:timelag-maps}) show negative 94-335 \AA{} time lags of $\approx-2000$ s in the longer loops on the lower edge of the \AR{}.} \added{However, in other parts of the \AR{},} \citeauthor{viall_survey_2017} found far fewer positive \added{and far more zero} 94-335 \AA{} time lags \added{compared to the simulated 94-335 \AA{} time lag maps for any of our heating scenarios}.

We also find negative 211-131 \AA{} time lags in the center of the \AR{} for the high-, intermediate-, and low-frequency cases, indicative of plasma cooling from the hot part of the 131 \AA{} channel through the 211 \AA{} channel. Though we have not shown them here, similar negative time lag signatures are present in nearly all of the other 131 \AA{} channel pairs as well. These results are consistent with that of \citet{cadavid_heating_2014} who found that in inter-moss regions of \AR{} NOAA 11250, intensity variations in the 131 \AA{} channel preceded brightenings in all other EUV channels. In the two control cases, we do not find any negative time lags as the cross-correlations in the core are dominated by uninterrupted cooling from 211 \AA{} to the cool part of 131 \AA{}.

For the 193-171 \AA{} channel pair, we find very few negative time lags because, unlike the 94 and 131 \AA{} channels, the 193 and 171 \AA{} channels are strongly peaked about a single temperature. Along with 211 \AA{}, these channels which are strongly-peaked about a single temperature are important for disambiguating the signals in channels with a bimodal temperature response function (see \autoref{fig:aia-response}).

\added{In all channel pairs,} our simulated time lags show far fewer zero time lags than the observations of \citet{viall_evidence_2012,viall_survey_2017} and the modeling work of \citet{bradshaw_patterns_2016} due to the lack of transition region emission in our model. \added{\citet{viall_transition_2015} found that many of the observed zero time lags were not present when observing an \AR{} off the limb, implying that most zero time lags are due to transition region dynamics.} Transition region emission shows near-zero time lag because every layer (or temperature) of the transition region responds in unison. However, for the 193-171 \AA{} channel pair, we find that zero time lags still dominate the inner core of the \AR{} for all five heating scenarios, suggesting that the plasma is cooling into, but not through the 171 \AA{} temperature bandpass \citep{viall_survey_2017}. This underscores the point that zero time lags do not imply steady heating \citep{viall_transition_2015,viall_signatures_2016}.

\subsubsection{Cross-Correlation Maps}\label{cross_correlation_maps}

% spell-checker: disable %

\begin{figure*}
	\centering
   \includegraphics[width=\columnwidth]{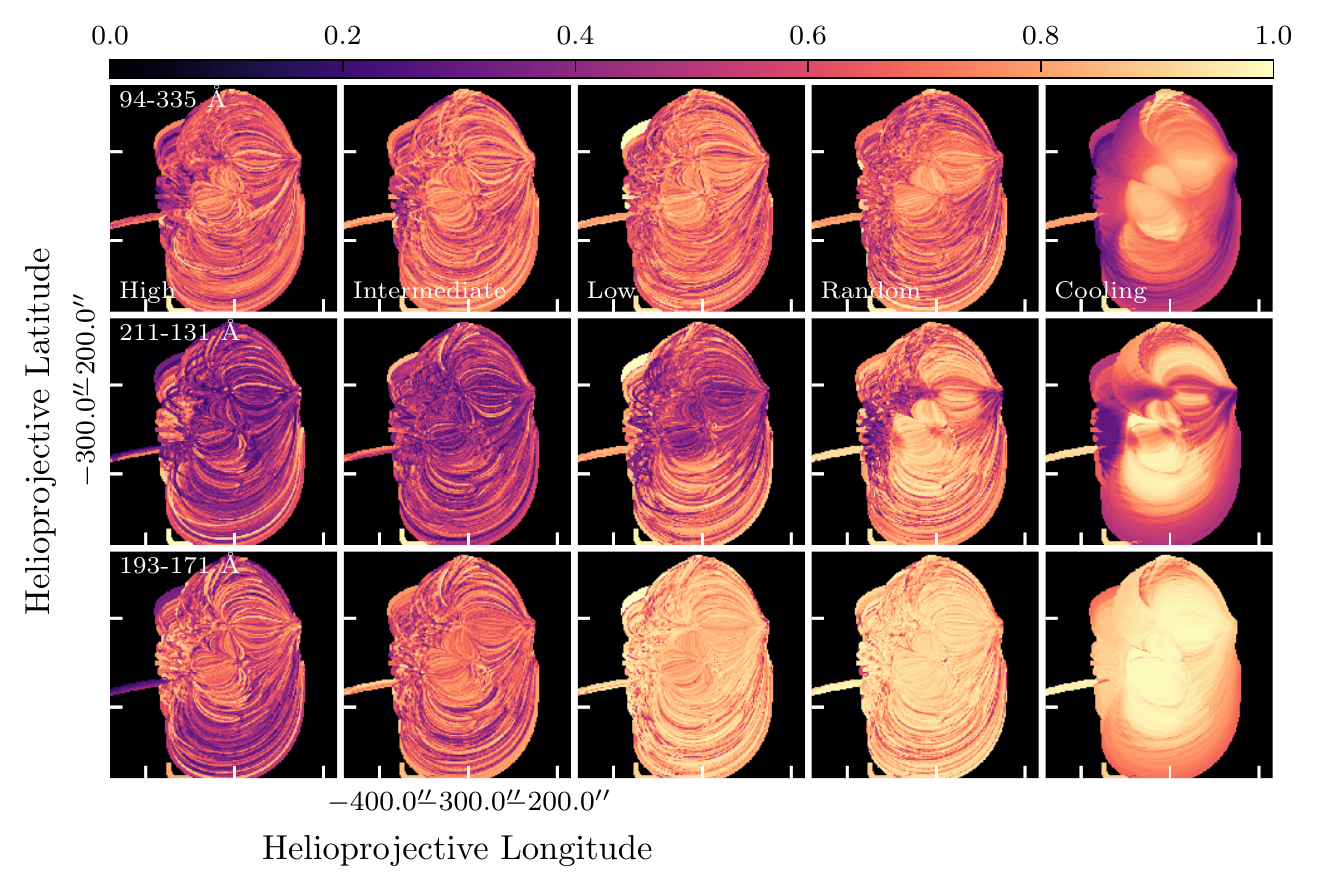}
	\caption{Same as \autoref{fig:timelag-maps} except each pixel shows the maximum cross-correlation, $\max\mathcal{C}_{AB}$ (see \autoref{eq:cc_pre}).}
	\label{fig:correlation-maps}
\end{figure*}
% spell-checker: enable %

\autoref{fig:correlation-maps} shows the peak cross-correlation value, $\max\mathcal{C}_{AB}$, for each selected channel pair. Looking first at all three channel pairs, we see that, on average, the cross-correlation increases as the heating frequency decreases. Additionally, we find that the highest cross-correlations tend to be in the center of the \AR{} while the lowest tend to be on the outer edge. Comparing \autoref{fig:correlation-maps} with the time lags in \autoref{fig:timelag-maps} also reveals that negative time lags are correlated with lower peak cross-correlation values. Furthermore, other than the ``cooling'' scenario, we find that there are large variations from one loop to the next for all heating frequencies such that the spatial coherence of these peak cross-correlation values is low. In \citetalias{barnes_understanding_2019-1}, we will use the peak cross-correlation value, in addition to the time lag, to classify the heating frequency in each observed pixel.

\subsubsection{Histograms}\label{histograms}

% spell-checker: disable %

\begin{figure*}
	\centering
   \includegraphics[width=\columnwidth]{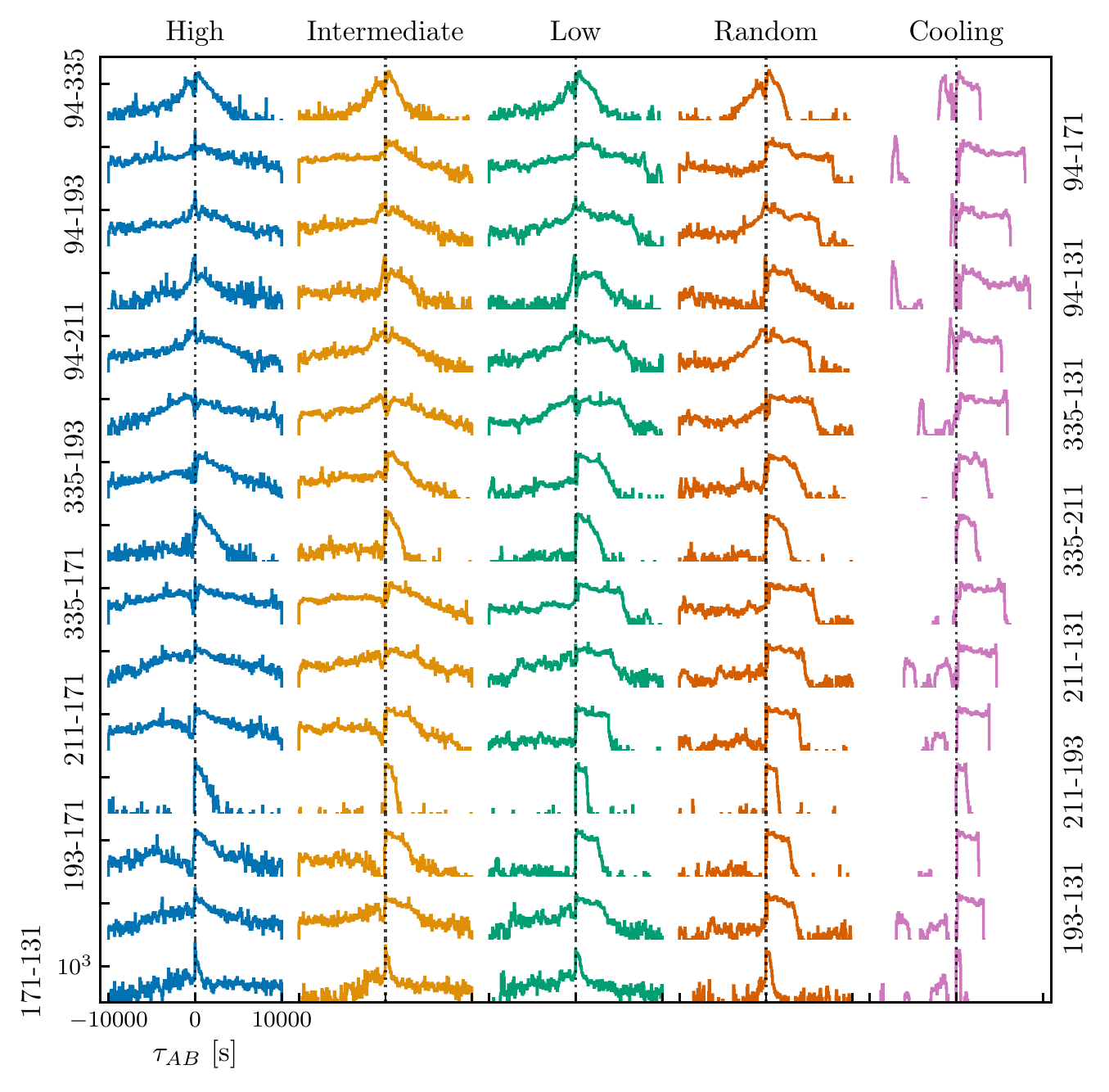}
	\caption{Histograms of time lags across the whole \AR{}. The rows indicate the different channel pairs and the columns indicate the different heating scenarios. Colors are used to denote the various heating scenarios. The black dashed line denotes zero time lag. The bin range is $\pm10^4$ s and the bin width is 60 s. As with the time lag maps, we do not include time lags corresponding to $\max{\mathcal{C}_{AB}}<0.1$.}
	\label{fig:timelag-histograms}
\end{figure*}
% spell-checker: enable %

\autoref{fig:timelag-histograms} shows histograms of time lags for every channel pair and all five heating scenarios. The time lags are binned between $-10^4$ s and $+10^4$ s in 60 s bins. Each histogram is colored according the corresponding heating scenario, consistent with \autoref{fig:hydro-profiles} and \autoref{fig:em-slope-histograms}. The columns are arranged such that heating frequency decreases from left to right. We show each channel pair for all heating models to demonstrate how the distribution of time lags evolves as the heating frequency varies.

Note that as the heating frequency decreases (from left to right), the number of negative time lags decreases. In the ``cooling'' case, there are very few negative time lags except for channel pairs which include one or both of the double-peaked channels (94 \AA{} and 131 \AA{}). For those channel pairs which include 94 \AA{} and/or 131 \AA{}, we expect to find negative time lags, even in the single-nanoflare cooling case as our convention of ordering the ``hot'' channel first has been violated such that cooling plasma can lead to negative time lags. For the remaining channel pairs, negative time lags are associated with the heating and cooling cycle being interrupted by repeated events on a given strand.
 %
%%%%%%%%%%%%%%%%%%%%%%%%%%%%%%%%%%%%%%%%%%%%%%%%%%%%%%%%%%%%%%%%%%%%%%%%%%%%%%%
%                               Discussion                                    %
%%%%%%%%%%%%%%%%%%%%%%%%%%%%%%%%%%%%%%%%%%%%%%%%%%%%%%%%%%%%%%%%%%%%%%%%%%%%%%%

\section{Discussion}\label{discussion}

For all of our heating models, we find negative time lags in at least one of the three channel pairs as shown in \autoref{fig:timelag-maps}. Negative time lags can be used to disambiguate the temperature sensitivity of the AIA passbands and can be produced in one of two ways: high-frequency heating in which the time lag is dominated by many frequent reheatings or a channel pair in which one channel is bimodal. While intensity in the 131 \AA{} channel can correspond to either $<0.4$ MK or $>10$ MK plasma (see \autoref{fig:aia-response}), negative time lags in the 211-131 \AA{} channel pair provide a possible signature of $\ge10$ MK plasma because a negative time lag implies the plasma is cooling from 131 \AA{} to 211 \AA{}. This also holds for the 171-131 and 193-131 \AA{} channel pairs as well while the 94-131 and 335-131 \AA{} channels are more ambiguous due to the first channel in the pair being bimodal as well.

As noted in \autoref{timelag_maps}, the high-, intermediate-, and low-frequency maps for the 211-131 \AA{} channel pair all show coherent negative time lags in the core. \added{We find far more negative 131 \AA{} time lags here compared to the observed 131 \AA{} time lags of \citep{viall_survey_2017} for this same \AR{}.} Because these strands are rooted in areas of strong field, enough energy is made available by the field (see \autoref{heating}) to heat them well into (and likely above) the hot component of the 131 \AA{} passband. Since these strands are relatively short, the density increases rapidly enough for this hot plasma to be visible before it is washed out by thermal conduction. \added{However, given that we are likely not overestimating the total energy budget of the \AR{} (see \autoref{em_distributions}), the excess of negative 131 \AA{} time lags seems to imply that the issue may be in the energy distribution. We plan to explore alternative heating scenarios in a future paper.}

Plasma undergoing pure cooling by radiation and thermal conduction produces a predictable and well-understood time lag signature. However, complicated heating scenarios and LOS geometries are likely to make it more difficult to interpret observed time lag signatures. Consider the case of a single cooling strand such that the maximum allowed time lag for a given channel pair $AB$ is the amount of time it takes to cool from $T_A$ to $T_B$ by thermal conduction and radiation. We may regard the ``cooling'' case in \autoref{fig:timelag-histograms} as the baseline time lag distribution given that all strands were heated only once at $t=0$ s. Because the time lag is primarily determined by the cooling phase of the strand, the time lag becomes primarily a function of the loop length $L$ since $\tau_\textup{cool}\propto L$. Two effects are likely to increase the decoherence of the baseline time lag distribution: multiple structures evolving out-of-phase along a given LOS (the ``random'' heating scenario) and multiple reheatings before the end of the cooling and draining cycle on a given strand. We note that multiple polluting structures along the LOS seem to primarily add negative time lags to the distribution (the ``random'' case) while increasing the frequency of events on a given strand extends the distribution in the positive direction. The latter effect also produces more negative time lags. Since steady heating can be thought of as nanoflare heating in the high-frequency limit ($\langle\twait\rangle\to0$), we expect the distribution of time lags to approach a uniform distribution as the heating frequency increases \added{because variations in the intensity will be increasingly dominated by noise} \citep{viall_signatures_2016}. \deleted{who found that} \added{In other words,} steadily-heated loops have no preferred time lag.

While our model for the energy deposition (see \autoref{heating}) does not assume any specific physical heating mechanism, the parameterization of the heating frequency in \autoref{eq:heating_types} has an interesting implication in the context of the \citet{parker_nanoflares_1988} nanoflare model. Rearranging \autoref{eq:heating_types} and recalling that $\tau_\textup{cool}\propto L$ gives $\langle\twait\rangle\propto L$, i.e. longer strands have longer absolute waiting times between heating events. Given that longer field lines tend to be rooted in regions of weaker magnetic field, this further implies that, where the field is stronger, energy is more quickly dissipated. According to \citet{parker_nanoflares_1988}, this dissipation is due to small-scale reconnection of flux tubes that are continually stressed by the convective motion of the underlying photosphere. Thus, in this context, our heating model implies that the reconnection and the underlying driver are more efficient in areas where the field is strongest.

Though we have not addressed it here, another possible mechanism for producing time-varying intensity in \AR s is thermal non-equilibrium (TNE) wherein condensation cycles driven by highly-stratified, but steady footpoint heating lead to long-period intensity pulsations \citep{kuin_thermal_1982}. Though originally used to explain coronal rain \citep{antolin_coronal_2010,antolin_multithermal_2015,auchere_coronal_2018} and prominences \citep{antiochos_model_1991}, several workers \citep{mok_three-dimensional_2016,winebarger_investigation_2016,froment_long-period_2017,winebarger_identifying_2018,froment_occurrence_2018} have recently claimed that TNE can produce time lag signatures similar to those of impulsive heating models, suggesting that observed time lags may be consistent with both impulsive and steady heating. However, it is not yet clear whether TNE is consistent with observed signatures of very hot (8-10 MK) plasma. Detailed comparisons between TNE and nanoflare simulations and observations are desperately needed.
 %
%%%%%%%%%%%%%%%%%%%%%%%%%%%%%%%%%%%%%%%%%%%%%%%%%%%%%%%%%%%%%%%%%%%%%%%%%%%%%%%
%                                   Summary and Conclusions                   %
%%%%%%%%%%%%%%%%%%%%%%%%%%%%%%%%%%%%%%%%%%%%%%%%%%%%%%%%%%%%%%%%%%%%%%%%%%%%%%%

\section{Summary}\label{conclusions}

We have carried out a series of numerical simulations in an effort to understand how signatures of the nanoflare heating frequency are manifested in two observables: the emission measure slope and the time lag. Additionally, we described each component of our pipeline for forward modeling \AR{} emission. For a given magnetogram observation of the relevant \AR{} (in this case, NOAA 1158), we compute a potential field extrapolation and trace a large number of field lines through the extrapolated vector field. For each traced field line, we run an EBTEL hydrodynamic simulation and use the resulting temperatures and densities, combined with data from CHIANTI and the instrument response function, to compute the time-dependent intensity. These intensities are then mapped back to the magnetic skeleton and integrated along the LOS in each pixel to create time-dependent images of the \AR{}.

Using our novel and efficient forward modeling pipeline, we produced AIA images for all six EUV channels for $\approx8$ hours of simulation time. From these results, we computed both the emission measure slope and the time lag for all possible channel pairs. We carried out these steps for three different nanoflare heating frequencies, high, intermediate, and low, (see \autoref{eq:heating_types}) in addition to two control models, for a total of five different heating scenarios (see \autoref{tab:heating}).

Our results can be summarized in the following points:
\begin{enumerate}
    \item As the heating frequency decreases, the emission measure slope, $a$, becomes increasingly shallow, saturating at $a\approx2$. As the heating frequency increases, the distribution of slopes over the \AR{} is shifted to higher values and broadens.
    \item The time lag becomes increasingly spatially coherent with decreasing heating frequency. When strands are allowed to cool without being re-energized, the spatial distribution of time lags is largely determined by the distribution of loop lengths over the \AR{}.
    \item The distribution of time lags becomes increasingly broad \deleted{and approaches a uniform distribution} as the heating frequency increases, consistent with the results of \citet{viall_signatures_2016}.
    \item Negative time lags in channel pairs where the second (``cool'') channel is 131 \AA{} provide a possible diagnostic for $\ge10$ MK plasma
\end{enumerate}

In this paper, we have used our advanced forward modeling pipeline to systematically examine how the emission measure slope and time lag are affected by the nanoflare heating frequency. In \citetalias{barnes_understanding_2019-1}, we use the model results presented here to train a random forest classifier and apply it to emission measure slopes and time lags derived from real AIA observations of NOAA 1158. The 15 channel pairs for the time lag and cross-correlation combined with the emission measure slope represent a 31-dimensional feature space and a single 500-by-500 pixel \AR{} amounts to $2.5\times10^5$ sample points. Performing an accurate assessment over this amount of data manually or ``by eye'' is at least impractical and likely impossible. Thus, the application of machine learning to the problem of assessing models in the context of real data is a critical step in understanding the underlying energy deposition in \AR{} cores and, to our knowledge, has not yet been applied in this context.  
 
%%%%%%%%%%%%%%%%%%%%%%%%%%%%%%%%%%%%%%%%%%%%%%%%%%%%%%%%%%%%%%%%%%%%%%%%%%%%%%%
%                                   Acknowledgment                            %
%%%%%%%%%%%%%%%%%%%%%%%%%%%%%%%%%%%%%%%%%%%%%%%%%%%%%%%%%%%%%%%%%%%%%%%%%%%%%%%
\acknowledgments
CHIANTI is a collaborative project involving George Mason University (USA), the University of Michigan (USA), University of Cambridge (UK), and NASA Goddard Space Flight Center (USA). This research makes use of SunPy, an open-source and free community-developed solar data analysis package written in Python \citep{sunpy_community_sunpypython_2015} and PlasmaPy, a community-developed open source core Python package for plasma physics \citep{plasmapy_community_2018_1238132}. SJB and WTB were supported by the NSF through CAREER award AGS-1450230. The work of NMV was supported by the NASA Supporting Research program. The complete source of this paper, including the data, code, and instructions for running the forward modeling code, can be found at \href{https://github.com/rice-solar-physics/synthetic-observables-paper-models}{github.com/rice-solar-physics/synthetic-observables-paper-models}.

%%%%%%%%%%%%%%%%%%%%%%%%%%%%%%%%%%%%%%%%%%%%%%%%%%%%%%%%%%%%%%%%%%%%%%%%%%%%%%%
%                                   Facilities                                %
%%%%%%%%%%%%%%%%%%%%%%%%%%%%%%%%%%%%%%%%%%%%%%%%%%%%%%%%%%%%%%%%%%%%%%%%%%%%%%%
\facility{SDO(AIA,HMI)}

%%%%%%%%%%%%%%%%%%%%%%%%%%%%%%%%%%%%%%%%%%%%%%%%%%%%%%%%%%%%%%%%%%%%%%%%%%%%%%%
%                                   Software                                  %
%%%%%%%%%%%%%%%%%%%%%%%%%%%%%%%%%%%%%%%%%%%%%%%%%%%%%%%%%%%%%%%%%%%%%%%%%%%%%%%
\software{
    Astropy \citep{the_astropy_collaboration_astropy_2018},
	Dask\citep{rocklin_dask_2015},
	IPython\citep{perez_ipython_2007},
	matplotlib\citep{hunter_matplotlib_2007},
	Numba\citep{lam_numba_2015},
	NumPy\citep{oliphant_guide_2006},
	PlasmaPy\citep{plasmapy_community_2018_1238132},
	Python\TeX\citep{poore_pythontex_2015},
	seaborn\citep{michael_waskom_2018_1313201},
	scipy\citep{jones_scipy_2001},
	SolarSoftware\citep{freeland_data_1998},
    SunPy\citep{stuart_mumford_2018_2155946},
    yt\citep{turk_yt_2011}
}

%%%%%%%%%%%%%%%%%%%%%%%%%%%%%%%%%%%%%%%%%%%%%%%%%%%%%%%%%%%%%%%%%%%%%%%%%%%%%%%
%                                   Appendix                                  %
%%%%%%%%%%%%%%%%%%%%%%%%%%%%%%%%%%%%%%%%%%%%%%%%%%%%%%%%%%%%%%%%%%%%%%%%%%%%%%%
\appendix
%
% spell-checker: disable %
% spell-checker: enable %
%%%%%%%%%%%%%%%%%%%%%%%%%%%%%%%%%%%%%%%%%%%%%%%%%%%%%%%%%%%%%%%%%%%%%%%%%%%%%%%
%                                   Appendix 1                                %
%%%%%%%%%%%%%%%%%%%%%%%%%%%%%%%%%%%%%%%%%%%%%%%%%%%%%%%%%%%%%%%%%%%%%%%%%%%%%%%
\section{Non-equilibrium Ion Populations}\label{nei}

In order to account for ionization non-equilibrium, we compute $f_{X,k}$ as a function of time $t$ by solving the time-dependent ion population equations for each ion $k$ in each element $X$,
\begin{equation}\label{eq:nei}
    \frac{\partial f_k}{\partial t} = n_e(R_{k+1}f_{k+1} + I_{k-1}f_{k-1} - I_kf_k - R_kf_k)
\end{equation}
where $n_e$ is the electron density and $R_k$ and $I_k$ are the temperature-dependent recombination and ionization rates of ion $k$, respectively. The ionization and recombination rates are computed using the data provided in CHIANTI. The ionization rates include both direct ionization and excitation autoionization and the recombination rates include both radiative and dielectronic recombination \citep[see section 6 of][]{young_chianti_2016}. Setting the left hand side of \autoref{eq:nei} to zero gives the equation of ionization equilibrium.

Note that for an element with atomic number $Z$, we must solve $Z+1$ coupled differential equations to find the non-equilibrium level populations. Following the approaches of \citet{masai_x-ray_1984,hughes_self-consistent_1985,shen_lagrangian_2015}, we can write \autoref{eq:nei} in matrix form,
\begin{equation}\label{eq:nei_mat}
    \frac{\partial}{\partial t}\mathbf{F} = \mathbf{A}\mathbf{F},
\end{equation}
where $\mathbf{F}=(f_1,f_2,\ldots,f_k,\ldots,f_{Z+1})$ and $\mathbf{A}$ is a ${Z+1\times Z+1}$ tridiagonal matrix containing the ionization and recombination rates, multiplied by the electron density,
\begin{equation}\label{eq:rate_mat}
    \mathbf{A} = n_e
        \begin{pmatrix}
            -I_1 & R_2 & 0 & \dots & 0 \\
            I_1 & -(I_2 + R_2) & R_3 & \dots & 0 \\
             & \ddots & \ddots & &  \\
            \vdots & I_{k-1} & -(I_k + R_k) & R_{k+1} & \vdots \\
             & & \ddots & \ddots & \\
            0 & \dots & 0 & I_{Z} & -R_{Z+1} 
        \end{pmatrix}.
\end{equation}

Due to drastic changes in the ionization and recombination rates with temperature, the above system of equations is very ``stiff,'' making explicit schemes extremely sensitive to the choice of time step \citep{macneice_numerical_1984,bradshaw_numerical_2009}. To solve \autoref{eq:nei_mat}, we use the ``deferred correction'' method of \citet{npl_modern_1961}, as pointed out by \citet{macneice_numerical_1984},
\begin{equation*}
    \mathbf{F}_{j+1} = \mathbf{F}_j + \frac{\Delta t}{2}\left(\frac{\partial}{\partial t}\mathbf{F}_{j+1} + \frac{\partial}{\partial t}\mathbf{F}_j\right) + \mathcal{O}(\Delta t^2),
\end{equation*}
where $\Delta t$ is the time step, $j$ indexes time, and $\mathcal{O}(\Delta t^2)$ denotes terms of second order or higher in $\Delta t$. Dropping the higher-order terms and using \autoref{eq:nei_mat} yields an expression for $\mathbf{F}_{j+1}$,
\begin{equation}\label{eq:nei_iterative}
    \mathbf{F}_{j+1} \approx \left(\mathbb{I} - \frac{\Delta t}{2}\mathbf{A}_{j+1}\right)^{-1}\left(\mathbb{I} + \frac{\Delta t}{2}\mathbf{A}_{j}\right)\mathbf{F}_j,
\end{equation}
where $\mathbb{I}$ is the identity matrix. To solve \autoref{eq:nei_mat}, we need only compute $\mathbf{A}_j$ for each $T(t_j)$, set $\mathbf{F}_0$ to the equilibrium ion populations, and iteratively compute \autoref{eq:nei_iterative} for all $j$. We solve \autoref{eq:nei_mat} for all elements in \autoref{tab:elements} and for each strand in the \AR{}.

Though this method is unconditionally stable, $\Delta t$ should still be chosen carefully as $\mathbf{F}$ will relax to the equilibrium charge states for very long time steps. In general, smaller time steps should be chosen when the electron temperature varies rapidly and the electron density is high. In this paper, we exploit the adaptive time step provided by the two-fluid EBTEL code which accounts for changes in the temperature and electron density. \citet{macneice_numerical_1984} provide two rules for adaptively adjusting the time step to ensure $f_k$ changes sufficiently slowly. Additionally, \citet{shen_lagrangian_2015} provide an alternate scheme for choosing the time step \textit{a priori} based on the input electron density and the eigenvalues of \autoref{eq:rate_mat}.

%%%%%%%%%%%%%%%%%%%%%%%%%%%%%%%%%%%%%%%%%%%%%%%%%%%%%%%%%%%%%%%%%%%%%%%%%%%%%%%%
%%                                   Appendix 2                                %
%%%%%%%%%%%%%%%%%%%%%%%%%%%%%%%%%%%%%%%%%%%%%%%%%%%%%%%%%%%%%%%%%%%%%%%%%%%%%%%%
\section{Effective AIA Response Functions}\label{effective_response_functions}

% spell-checker: disable %

\begin{figure*}
	\centering
   \includegraphics[width=\columnwidth]{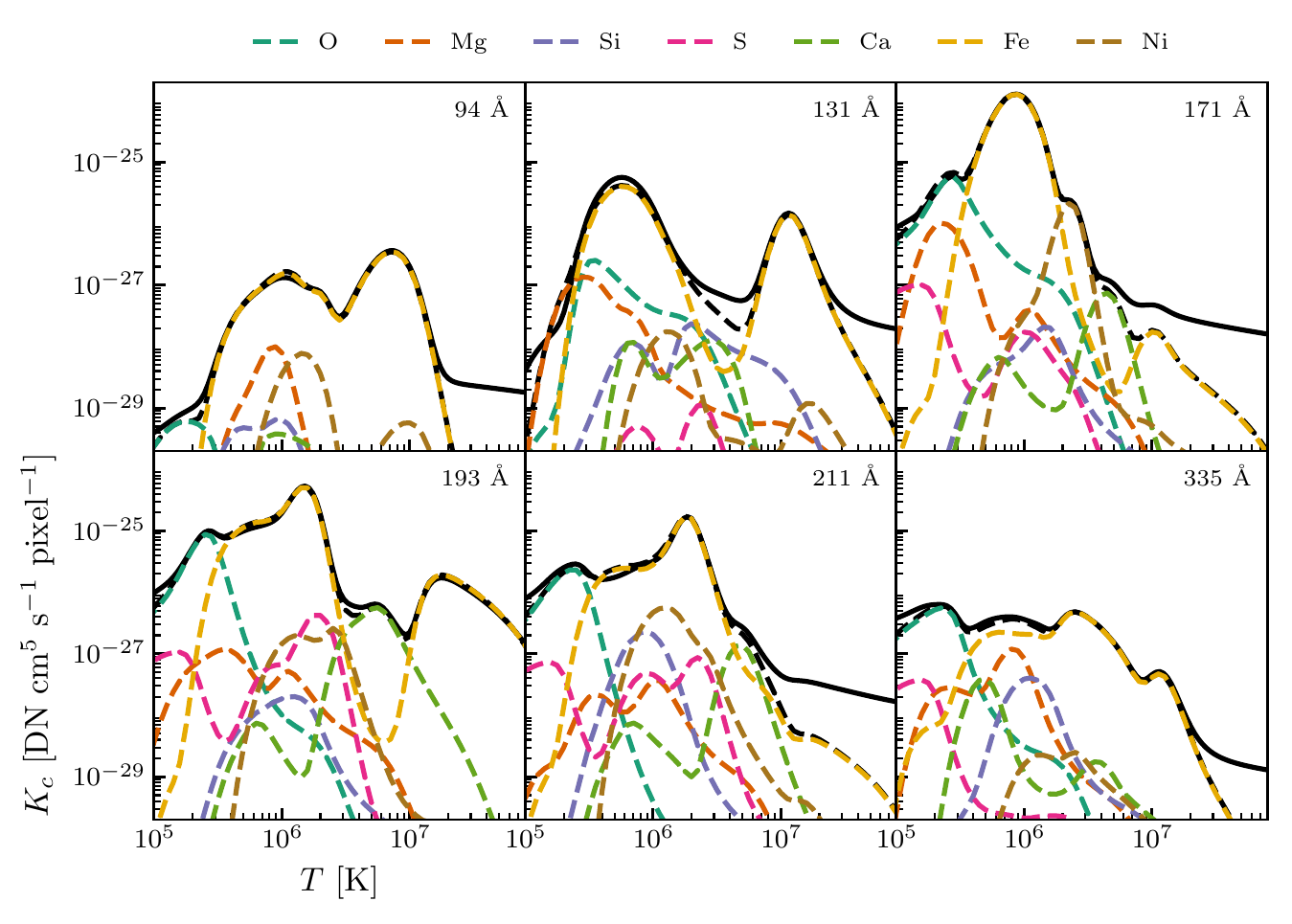}
	\caption{SSW temperature response functions (solid black) and effective temperature response functions for the elements in \autoref{tab:elements} (dashed black) for all six EUV AIA channels. The colored, dashed curves, as indicated in the legend, denote the contributions of the individual elements to the total response. For this calculation, we have assumed equilibrium ionization and a constant pressure of $10^{15}$ K cm$^{-3}$. We do not account for the time-varying degradation of the instrument.}
	\label{fig:aia-response}
\end{figure*}
% spell-checker: enable %

As discussed in \autoref{atomic}, the assumption of ionization equilibrium is likely to be violated in the impulsive heating cases considered here. Thus, we must recompute the contributions of each ion to the total channel response, using the result of \autoref{eq:nei_mat} in place of the equilibrium ion population fractions. \autoref{fig:aia-response} shows the effective temperature response functions for the six EUV channels on AIA compared to those calculated from \texttt{aia\_get\_response.pro} in SSW. Even though we include a limited number of transitions from the CHIANTI database (see \autoref{tab:elements}), we recover nearly all of the response from each channel. The high-temperature contributions in the SSW functions are due to continuum emission which we do not include in our model. In all cases, the continuum contribution is several orders of magnitude below peak of the channel response. Additionally, we do not account for the time variation in the wavelength response functions due to the degradation of the detector \citep[see Section 2.1.6 of ][]{boerner_initial_2012}.

% spell-checker: disable %

\begin{figure}
	\centering
   \includegraphics[width=0.5\columnwidth]{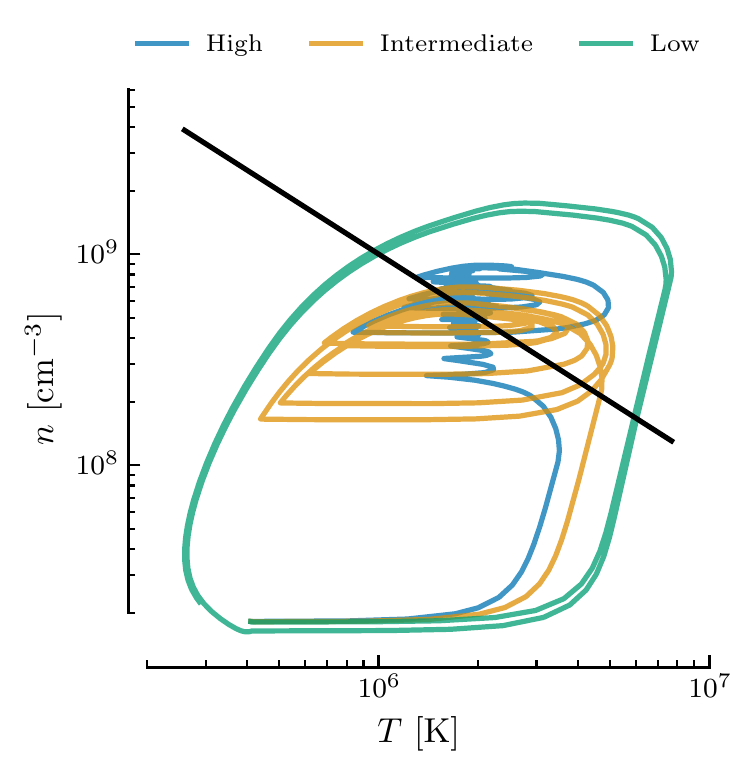}
	\caption{$n-T$ phase-space orbits for a single strand for the three heating scenarios as defined by the legend. The black line indicates a constant pressure of $10^{15}$ K cm$^{-3}$.}
	\label{fig:hydro-phase-space}
\end{figure}
% spell-checker: enable %

Furthermore, during the evolution of a strand, the pressure is not constant for any of our heating scenarios as evidenced by \autoref{fig:hydro-phase-space}. The black line of constant pressure $p=10^{15}$ K cm$^{-3}$ shows the default pressure at which the SSW AIA response functions are evaluated. The other lines show the temperature-density phase space evolution for the high-, intermediate-, and low-frequency cases for a single strand, none of which is well described by the assumption of constant pressure. By recomputing and interpolating the emissivity to the temperatures and densities as defined by our hydrodynamic simulation, we ensure that we are evaluating all quantities in \autoref{eq:ppuv} at the correct temperature and density.

%%%%%%%%%%%%%%%%%%%%%%%%%%%%%%%%%%%%%%%%%%%%%%%%%%%%%%%%%%%%%%%%%%%%%%%%%%%%%%%
%                                   Appendix 3                                %
%%%%%%%%%%%%%%%%%%%%%%%%%%%%%%%%%%%%%%%%%%%%%%%%%%%%%%%%%%%%%%%%%%%%%%%%%%%%%%%
\section{Computing Time Lags}\label{timelag_details}

To find the associated time lag for a channel pair in a given pixel, we compute the cross-correlation between the time series associated with each channel and find the delay which maximizes this cross-correlation. We can express the cross-correlation $\mathcal{C}$ between two channels $A$ and $B$ as,
\begin{equation}\label{eq:cc_pre}
    \mathcal{C}_{AB}(\tau) = \mathcal{I}_A(t)\star\mathcal{I}_B(t) = \mathcal{I}_A(-t)\ast\mathcal{I}_B(t)
\end{equation}
where $\star$ and $\ast$ represent the correlation and convolution operators, respectively, $\tau$ is the lag and
\begin{equation*}
    \mathcal{I}_c(t)=\frac{I_c(t)-\bar{I}_c}{\sigma_{c}},
\end{equation*}
is the intensity of channel $c$ as a function of time, $I(t)$, centered to zero mean and scaled to unit standard deviation. Taking the Fourier transform, $\mathscr{F}$, of both sides of \autoref{eq:cc_pre}, using the convolution theorem, and then taking the inverse Fourier transform, $\mathscr{F}^{-1}$, gives,
\begin{align}\label{eq:cc}
    \fourier{\mathcal{C}_{AB}(\tau)} &= \fourier{\mathcal{I}_A(-t)\ast\mathcal{I}_B(t)}, \nonumber\\
    &= \fourier{\mathcal{I}_A(-t)}\fourier{\mathcal{I}_B(t)}, \nonumber\\
    \mathcal{C}_{AB}(\tau) &= \inversefourier{\fourier{\mathcal{I}_A(-t)}\fourier{\mathcal{I}_B(t)}}.
\end{align}
Scaling \autoref{eq:cc} by the length of the intensity time series $I(t)$ yields the same result as that of the correlation defined in Section 2 of \citet{viall_evidence_2012}. Furthermore, the \textit{time lag} between channels $A$ and $B$ is defined as,
\begin{equation}\label{eq:timelag}
    \tau_{AB} = \argmax_{\tau}\,\mathcal{C}_{AB}(\tau).
\end{equation}

By exploiting the definition of the cross-correlation as given in \autoref{eq:cc}, we can leverage existing Fourier transform algorithms in order to compute $\mathcal{C}_{AB}$ in a scalable and efficient manner. For a 500-by-500 pixel active region observation and 15 possible channel pairs, we need to compute $\tau_{AB}$ nearly $4\times10^6$ times. We use the highly-optimized and thoroughly tested Fourier transform algorithms in the NumPy library for array computations \citep{oliphant_guide_2006} combined with the Dask library for parallel and distributed computing \citep{rocklin_dask_2015}. Using Dask, we are able to parallelize this operation over many cores such that, on a 64-core machine, we can compute time lags for all 15 channel pairs in every pixel of the \AR{} in less than ten minutes. For comparison, doing the same calculation by calling the IDL function \texttt{c\_correlate.pro} on each pixel in serial would take $\approx14$ hours for all 15 channel pairs.
 
%%%%%%%%%%%%%%%%%%%%%%%%%%%%%%%%%%%%%%%%%%%%%%%%%%%%%%%%%%%%%%%%%%%%%%%%%%%%%%%
%                                   References                                %
%%%%%%%%%%%%%%%%%%%%%%%%%%%%%%%%%%%%%%%%%%%%%%%%%%%%%%%%%%%%%%%%%%%%%%%%%%%%%%%
\bibliographystyle{aasjournal.bst}
\bibliography{references.bib,software.bib}

\listofchanges
\end{document}